\documentclass[11pt,a4paper]{iopart}
\pdfoutput=1
\usepackage{cite}
\usepackage{graphicx} 
\usepackage{xcolor}

\usepackage[utf8]{inputenc}

\usepackage{iopams}

\expandafter\let\csname equation*\endcsname\relax

\expandafter\let\csname endequation*\endcsname\relax 

\usepackage{amsmath}
\usepackage{amssymb}
\usepackage{hyperref}
\usepackage{enumitem}

\makeatletter
\def\@mkboth#1#2{}
\newlength\appendixwidth
\newcommand{\patchl@section}{%
  \settowidth{\appendixwidth}{\textbf{Appendix }}%
  \addtolength{\appendixwidth}{1.5em}%
  \patchcmd{\l@section}{1.5em}{\appendixwidth}{}{\ddt}%
}
\makeatother

\begin{document}

\title{Dynamically emergent  correlations in a Brownian gas with diffusing diffusivity}

\author{Nikhil Mesquita}
\address{Raman Research Institute, Bangalore 560080, India\\
Email: nikhilm@rrimail.rri.res.in}

\author{Satya N. Majumdar}
\address{LPTMS, CNRS, Universit\'e Paris-Sud, Universit\'e Paris-Saclay, 91405 Orsay, France\\
Email: satyanarayan.majumdar@cnrs.fr}

\author{Sanjib Sabhapandit}
\address{Raman Research Institute, Bangalore 560080, India\\
Email: sanjib@rri.res.in}

\begin{abstract}

We study a gas of $N$ Brownian particles in the presence of a common stochastic diffusivity $D(t)=B^2(t)$, where $B(t)$ represents a one-dimensional Brownian motion at time $t$. Starting from all the particles localized at the origin, the gas expands with a ballistic scaling $x\sim t$. We show that because of the common stochastic diffusivity, the expanding gas gets dynamically correlated, and the joint probability density function of the position of the particles has a conditionally independent and identically distributed (CIID) structure that was recently found in several other systems. The special CIID structure allows us to compute the average density profile of the gas, extreme and order statistics, gap distribution between successive particles, and the full counting statistics (FCS) that describes the probability density function (PDF) $H(\kappa, t)$ of the fraction of particles $\kappa$ in a given region $[-L,L]$. Interestingly, the position fluctuation of the central particles and the average density profiles are described by the same scaling function. The PDF describing the FCS has an essential singularity near $\kappa=0$, indicating the presence of particles inside the box $[-L,L]$ at all times. Near the upper limit $\kappa =1$, the scaling function $H(\kappa,t)$ has a rather unusual behavior: $H(\kappa,t)\sim (1-\kappa)^{\beta(t)}$ where the exponent $\beta(t)$ changes continuously with time. At early times $\beta(t)$ is negative, indicating a divergence of $H(\kappa,t)$ as $\kappa\to 1$, whereas  $\beta(t)$ becomes positive for $t>t_c$ where $t_c$ is computed exactly. For $t>t_c$, the scaling function $H(\kappa,t)$ vanishes as $\kappa\to 1$, indicating that it is highly unlikely to have all the particles in the interval $[-L,L]$. Exactly at $t=t_c$, $\beta=0$, indicating that the PDF approaches a non-zero constant as $\kappa\to 1$. Thus, as a function of $t$, the FCS exhibits an interesting shape transition.
We also obtain the PDFs of the first-passage time to a given position $x$ and first-exit time from a box $[-L,L]$, by any one of the particles, and find that both PDFs are described by the same scaling function.
\end{abstract}

\noindent\rule{\hsize}{2pt}
\tableofcontents
\noindent\rule{\hsize}{2pt}
\maketitle

\section{Introduction}
\label{sec:Intro}

A system of non-interacting particles may become dynamically correlated when the particles are coupled to a stochastically fluctuating environment. These correlations arise dynamically, and are not built in, i.e., they get generated even though there is no direct interaction between the particles. These correlations emerge because the particles share the same fluctuating environment. Such dynamically emerging correlations have recently been uncovered theoretically in a number of simple models~\cite{BLMS2023, BLMS2024, BKMS2024, SM2024}, and have also been demonstrated experimentally~\cite{cilberto}. In all the theoretical models studied so far, the fluctuating environment drives the system into a nonequilibrium stationary state (NESS) in which there is an all-to-all effective attraction between the particles, which makes them strongly correlated. This includes, e.g., $N$ non-interacting Brownian particles that are simultaneously reset to the origin at a constant rate~\cite{BLMS2023, BLMS2024}, and also $N$ particles in a harmonic trap, whose stiffness fluctuates between two values with rates $r_1$ and $r_2$~\cite{BKMS2024}, or the center of the trap undergoes a stochastic modulation~\cite{SM2024}. Such a strongly correlated NESS has also been found in a quantum system consisting of $N$ non-interacting bosons in a harmonic trap subjected to a specific quantum resetting protocol~\cite{KMS2025}.

However, in some systems, the fluctuating environment may not drive the system into a NESS, and the state of the system is always time-dependent. It is interesting to ask how the correlations between the particles grow with time in such time-dependent systems. In this paper, we present an exactly solvable model with such time-dependent emergent correlations created by the fluctuating environment. Our model is just an $N$-body generalization of a well-studied single Brownian particle subjected to a diffusing diffusivity. More precisely, we consider $N$ non-interacting Brownian particles on a line with coordinates $\{x_1(t), x_2 (t), \dotsc, x_N(t)\}$ that evolve as
\begin{equation}
    \frac{dx_i}{dt} = \sqrt{2 D(t)} \, \eta_i(t) , 
    \label{eq:LE}
\end{equation}
where $D(t) >0$ represents a fluctuating positive diffusivity, which itself performs an independent stochastic motion. The set $\{\eta_1(t), \eta_2(t), \dotsc,\eta_N(t)\}$ represents independent white noise with zero mean and a correlator 
$\langle \, \eta_j(t) \,  \eta_k(t') \, \rangle =\delta_{j,k} \, \delta(t-t')$.
The dynamics of $D(t)$ is not affected by the dynamics of $x_i(t)$'s. However, the dynamics of $x_i(t)$'s for each $i$ share the common fluctuating diffusivity $D(t)$, which makes them correlated. The dynamics of a single particle ($N=1$) for different choices of $D(t)>0$, such as the square of a one-dimensional Brownian motion or a one-dimensional Ornstein-Uhlenbeck (OU) process,  have been studied in the literature~\cite{CS14, chechkin2017brownian, tyagi2017non, lanoiselee2018model, SGMSS2020, barkai2020packets, pacheco2021large, hamdi2024laplace, singh2024emergence, gueneau2025large, hidalgo2021cusp,sposini2018first,jain2017diffusing,jain2016diffusingsurvival,yin2021non}. The above model of $N$ non-interacting Brownian particles with a common fluctuating diffusing diffusivity $D(t)$ is equivalent to an $N$-dimensional Brownian motion with diffusing diffusivity, for which the probability density function (PDF) of the radial coordinate $R=\sqrt{x_1^2+x_2^2+\dotsb+x_N^2}$ has been studied for  $D(t)=B^2(t)$~\cite{SBS2022}, where $B(t)$ represents a one-dimensional Brownian motion at time $t$, i.e., 
\begin{equation}
       \frac{dB}{dt} = \sqrt{2\Lambda^2}\,\zeta(t),\quad\text{with }~ B(0)=0.
    \label{eq:bm}
\end{equation}
Here $\Lambda>0$ is a constant that has the dimension of velocity and $\zeta(t)$ is a Gaussian white noise with zero mean and two-time correlation function $\langle  \zeta(t) \zeta(t')\rangle =\delta(t-t')$. 
The choice of stochastic diffusivity $D(t)=B^2(t)$ might model a medium undergoing heating where one expects the amplitude of the random force driving the Brownian particle typically increases with time, but in a fluctuating manner. It was found that  (i) the typical $R$ grows linearly with time and (ii) the PDF of the scaled distance $R/t$ has a universal exponential tail in all dimensions $N$, with the dependence on $N$ appearing only through the sub-leading prefactor~\cite{SBS2022}. However, in this paper, we are interested in observables such as the average density, the order statistics, the gap statistics, and the full counting statistics. These observables make sense only if we interpret $x_i(t)$'s as the position of $N$ independent particles on a line, rather than interpreting them as the components of the position of a single particle in $N$ dimensions. For example, the gap between two consecutive particles makes sense only in one dimension, since there is no physical meaning of the gap between two components of a single vector in $N$ dimensions.  
 
Let us also remark that another interesting recent example of diffusing diffusivity---specifically with the choice $D(t)=B^2(t)$---has been studied in the context of a Rouse polymer chain, or equivalently, a Gaussian interface evolving according to the Edwards–Wilkinson equation~\cite{DMS2025}.
It was found that the typical height at a fixed point in space grows as $t^{3/4}$ at late times. Moreover, the scaled height approaches a highly non-trivial non-Gaussian distribution that was computed exactly~\cite{DMS2025}. In this Rouse model, the monomers have pairwise harmonic interactions between them. The model that we consider here is even simpler. Unlike the interacting monomers in the Rouse chain, here the particles have no direct interaction between them. They only share the common diffusivity, which we choose to be $D(t)=B^2(t)$, as in the Rouse chain or the interface model above. Our goal in this paper is to see how the correlations between particles emerge dynamically from sharing the same time-dependent fluctuating diffusivity $D(t)=B^2(t)$ and how these correlations affect the behavior of different physical observables of this gas of particles. This is done exactly by computing the joint probability density function (JPDF) $P(x_1, x_2, \dotsc, x_N, t)$ at any time $t$. Furthermore, we show that this JPDF in our model at any time $t$ also has a \emph{conditionally independent and identically distributed} (CIID) structure found recently in a number of other models~\cite{BLMS2023, BLMS2024, BKMS2024, SM2024, KMS2025, biroli2025resetting}. This special CIID structure allows the exact computation of various macroscopic and microscopic observables. The average density profile $\rho(x,t)$ of the gas at time $t$ is an example of a macroscopic observable. Microscopic observables include the extreme and the order statistics, the gap statistics, and the full counting statistics. These observables have been computed recently for other models mentioned above with a CIID structure in the steady state~\cite{BLMS2023, BLMS2024, BKMS2024, SM2024, KMS2025, biroli2025resetting}. The difference in our case is that we do not have a steady state for the choice $D(t)=B^2(t)$, and hence these observables are time dependent. Nevertheless, exploiting the CIID structure, these observables can be computed exactly at all times $t$. Moreover, one can also analyze temporal properties, since we have an example of a strongly correlated time-dependent system. In particular, here, we study the first-passage properties of this strongly correlated model.

The rest of the paper is organized as follows. We first obtain the JPDF in \sref{s:JPDF} and show that it has the CIID structure. Using the JPDF, we compute the other observable in the subsequent sections. In \sref{s:cor}, we compute the correlations between the particles and show that there are strong positive correlations between the particles, characterized by the correlation function $\langle x_i^2 x_j^2 \rangle - \langle x_i^2 \rangle \langle x_j^2 \rangle$ that increases as $t^4$ with time. In fact, the gas in our model expands ballistically with time with a length-scale $\xi(t)\propto t$, and in \sref{s:density} we find the density profile of the gas by scaling the space with $\xi(t)$.   \Sref{s:order} contains the order statistics where we investigate the distribution of the $k$-th maximum. The two special cases $k=N/2$ and $k=1$ are treated more carefully in sub-sections \ref{s:central} and \ref{Section:EVS} respectively. In \sref{s:gap} we consider the gap statistics, and  \sref{s:FCS} contains the full counting statistics.  In \sref{sec:FPT}, we study first-passage properties, such as the survival probability, PDFs of the first-passage time to a given level $x$ and first-exit time from a given box $[-L,L]$. \Sref{sec:simulation} contains some details of numerical simulation. We conclude in \sref{s:conclusion}.

\section{The JPDF and the CIID structure}
\label{s:JPDF}

We start by computing the exact JPDF $P(x_1, x_2, \dotsc, x_N,t)$ of the positions of the particles at any time $t$. 
Assuming that $x_i(0)=0$ for all $i$, it follows from~\eref{eq:LE} that 
\begin{equation}
x_i(t) = \int_0^t\, \sqrt{2 D(\tau)}\, \eta_i(\tau)\, d\tau.
\label{eq:xt} 
\end{equation}
Hence, for any given realization of the fluctuating diffusivity process $\{D(\tau); 0\le \tau\le t\}$, the process $\{x_1(t), x_2(t), \dotsc, x_N(t)\}$ is an independent multivariate Gaussian process with a JPDF
\begin{equation}
P(x_1, x_2, \dotsc, x_N,t|\{D(\tau)\}) = \prod_{j=1}^N \frac{1}{\sqrt{2\pi V(t)}}\, \exp\left(-\frac{x_j^2}{2V(t)}\right),
\label{eq:JPDF|V} 
\end{equation}
where the variance $V(t)$ of $x_i$ for any $i$, for a given realization of $\{D(\tau); 0\le \tau\le t\}$, is given by the functional 
\begin{equation}
V(t) = 2\int_0^t D(\tau)\, d\tau \, .
\label{eq:defV0}
\end{equation} 
This is easily computed from~\eref{eq:xt} using the properties of the white noise, namely,  $\langle \, \eta_j(\tau) \rangle=0$ and 
\begin{math}
    \langle \, \eta_j(\tau) \,  \eta_k(\tau') \, \rangle =\delta_{j,k} \, \delta(\tau-\tau') .
\end{math} 
Note that the dependence of the JPDF in~\eref{eq:JPDF|V} on the realization of $\{D(\tau); 0\le \tau\le t\}$ appears only through $V(t)$, which is an integral over $2 D(\tau)$. Hence, averaging JPDF in~\eref{eq:JPDF|V} over the realization of the full process $\{D(\tau); 0\le \tau\le t\}$ amounts to averaging over the random variable $V(t)$ given in~\eref{eq:defV0}, i.e., 
\begin{equation}
P(x_1, x_2, \dotsc, x_N,t) = \int_0^\infty P(x_1, x_2, \dotsc, x_N,t|\{D(\tau)\}) \, h(V,t)\, dV\, ,
\end{equation}
where $h(V,t)$ denotes the PDF of the random variable $V$ at time $t$. 

With the choice $D(\tau)=B^2(\tau)$, the variance $V(t)$ in~\eref{eq:defV0} can be interpreted as a Brownian functional 
\begin{equation}
V(t) = 2\int_0^t B^2(\tau)\, d\tau,
\label{eq:defV}
\end{equation} 
where $B(\tau)$ is a Brownian motion defined in~\eref{eq:bm}. 
 Then, averaging~\eref{eq:JPDF|V} over $V$, the JPDF of the positions of the particles at any time $t$ is given by
 \begin{equation}
P(x_1, x_2, \dotsc, x_N,t) = \int_0^\infty\, dV\, h(V,t)\, \prod_{j=1}^N \frac{1}{\sqrt{2\pi V}}\, \exp\left(-\frac{x_j^2}{2V}\right),
\label{eq:JPDF} 
\end{equation}
where the only unknown so far is the PDF $h(V,t)$ of $V$ defined in~\eref{eq:defV}.

The PDF $h(V,t)$ of the functional of the square of a Brownian motion defined in~\eref{eq:defV} has long been known in the mathematics literature~\cite{cameron1944wiener, cameron1945transformations, EK1946, kac1949distributions} --- also see~\cite{SBS2021, SBS2022} for a physicist's derivation. Its Laplace transform, with $B(t)$ defined in~\eref{eq:bm}, is given by
\begin{equation}
\tilde{h}(\lambda,t)=\int_0^\infty e^{-\lambda V} \, h(V,t)\, dV =\sqrt{\mathrm{sech} \bigl(2  \Lambda t \, \sqrt{2 \lambda}\bigr)} ~.
\label{eq:LThV}
\end{equation}
Therefore, $h(V,t)$ is formally given by the Bromwich integral
\begin{equation}
h(V,t)= \frac{1}{2\pi i} \int_{\gamma_1- i \infty}^{\gamma_1+i \infty} \, d\lambda \,  e^{\lambda V}\,  \tilde{h}(\lambda,t) =  \frac{1}{2\pi i } \int_{\gamma_1- i \infty}^{\gamma_1+i \infty} \, d\lambda \, e^{\lambda V}\sqrt{\mathrm{sech}(2 \Lambda t \sqrt{2 \lambda})}~,
\label{eq:hV1}
\end{equation}
where $\gamma_1$ is a real constant such that all the singularities of the function $\tilde{h}(\lambda,t)$ lie on the left of the contour in the complex $\lambda$ plane. 
It follows from~\eref{eq:defV} that since typically $B(\tau)$ scales as $\sqrt{\tau}$, the quantity $V(t)$ scales as $t^2$ for any $t$. Consequently, one expects that $h(V,t)$ must have a scaling form $h(V,t)\sim t^{-2} q (V/t^2) $. Actually, it turns out to be convenient to write the PDF in the scaling form
\begin{equation}
 h(V,t) = \frac{1}{[\xi(t)]^2} \,  Q\left(\frac{V}{[\xi(t)]^2} \right) \ ,\quad\text{with}\quad 
 \xi(t) = \sqrt{2} \, \Lambda \, t,
 \label{eq:hV2} 
\end{equation}
where the scaling function  $Q(z)$ can be read off from~\eref{eq:hV1} as 
\begin{equation}
    Q(z) = \frac{1}{2\pi i} \int_{\gamma- i \infty}^{\gamma+i \infty} ds \, e^{sz}\, \sqrt{\mathrm{sech}( 2\sqrt{s})} \ . 
    \label{eq:Q1}
\end{equation}
Here, $\gamma$ is a real constant such that all the singularities of $\sqrt{\mathrm{sech}( 2\sqrt{s})}$ lie on the left of the contour in the complex $s$ plane. While it is hard to perform the integral in~\eref{eq:Q1} exactly, it can be easily analyzed to extract the small and large $z$ behaviors of the function $Q(z)$.

\begin{figure}
    \centering
\includegraphics[width=0.45\textwidth]{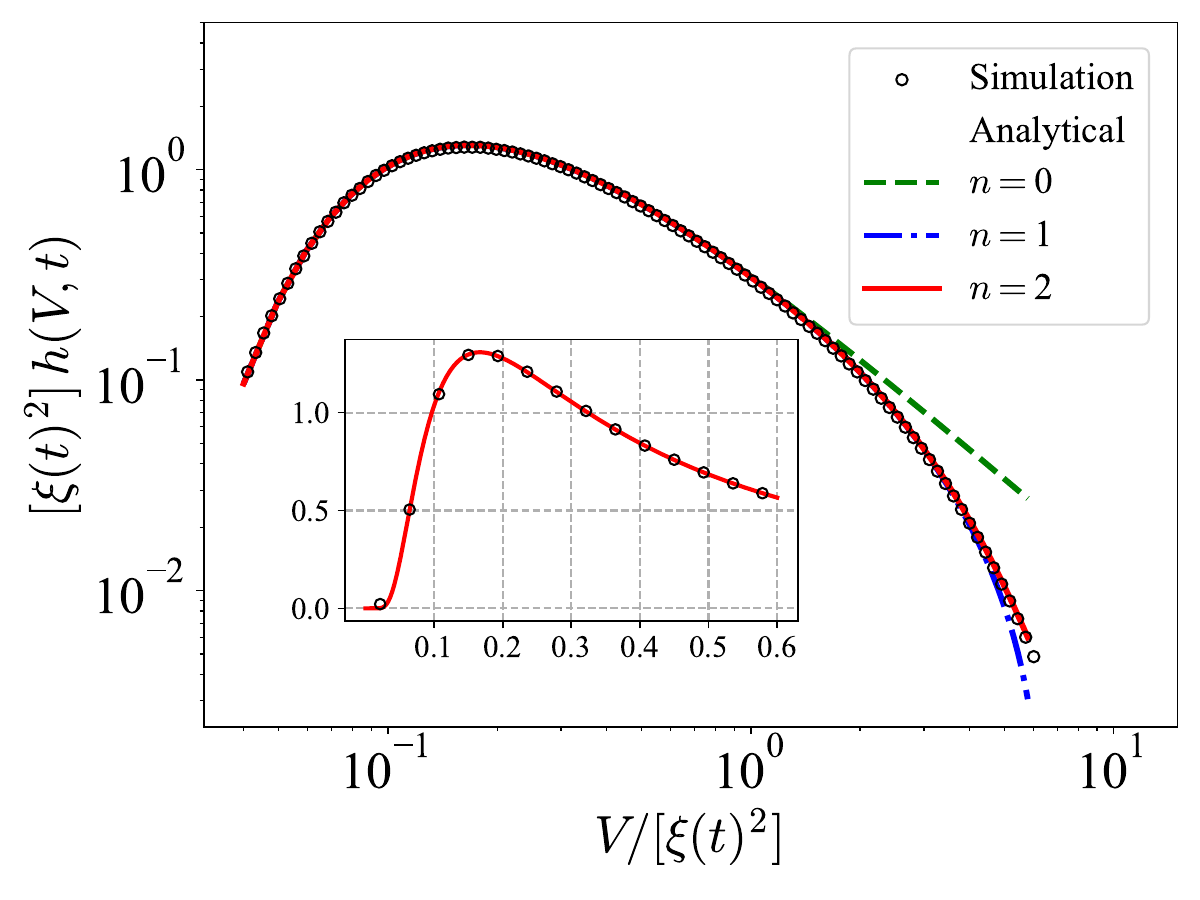}~
\includegraphics[width=0.45\textwidth]{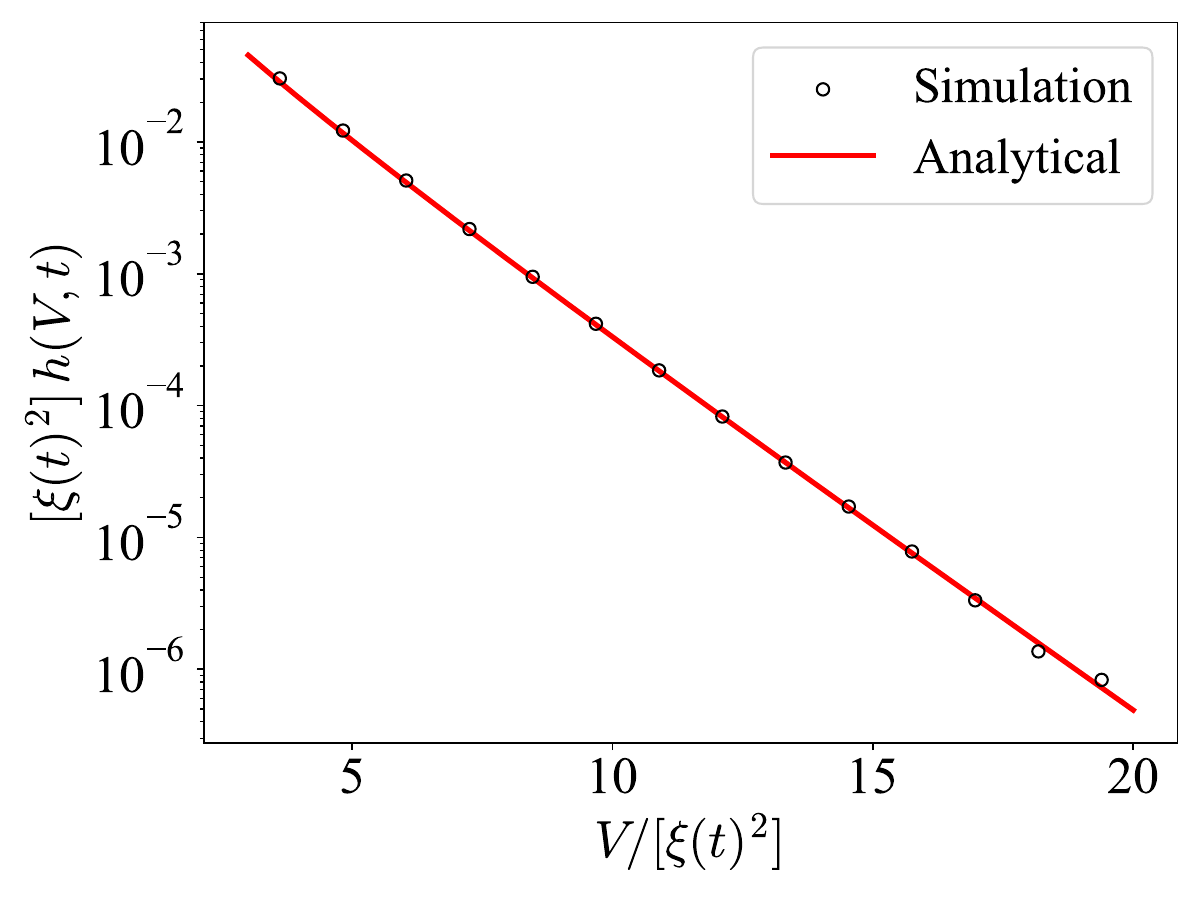} 
   \caption{The PDF $Q(V/[\xi(t)]^2) \equiv [\xi(t)]^2 \,  h(V,t)$  of the scaled variable $V/[\xi(t)]^2$.
   The points are from numerical simulations (detailed in \ref{sec:sim_h_V}) for the parameters $\Lambda=1$, $dt =0.01$, $t=10$, averaged over $ 10^8$ realizations. The left panel highlights the small $z$ behavior, while the right one highlights the tail behavior. Left: The solid lines are analytical plots for different truncations of the infinite series in equation~\eqref{Q(z)}. Right: The solid line plots the asymptotic tail given in~\eqref{eq:Q large z}.}
    \label{fig:h(V)}
\end{figure}

To obtain the small $z$ behavior, it is useful to express the $\sqrt{\mathrm{sech}( 2\sqrt{s})}=\sqrt{2}\, e^{-\sqrt{s}}/\sqrt{1+e^{-4\sqrt{s}}}$  in~\eref{eq:Q1}  and then expand the denominator using the binomial series, as 
\begin{equation}
    Q(z) =\sqrt{2} \sum_{n=0}^\infty  \, \binom{- 1/2}{n} \, \int_{\gamma- i \infty}^{\gamma+i \infty} \frac{ds}{2 \pi i} \,  e^{z s}  \, e^{ - (4n+1) \sqrt{s} } \ .
    \label{eq:Q2}
\end{equation}
The evaluation of the Bromwich integral in~\eref{eq:Q2} for any $n$ --- where the integrand has a branch-point at $s=0$ --- is quite standard. The inverse Laplace transform yields,
\begin{equation}
    Q(z) = \frac{1}{\sqrt{2 \pi}} \, \sum_{n=0}^\infty \binom{-\frac{1}{2}}{n} \, \frac{4n+1}{z^{3/2}} \, \exp{\left[-\frac{(4n+1)^2}{4 \, z }\right]} \, . 
    \label{Q(z)} 
\end{equation}
While the infinite series in~\eref{Q(z)} is an exact expression of $Q(z)$, it is mostly useful for describing the small $z$ behavior, by keeping only a few terms of the series, as shown in the left panel in~\fref{fig:h(V)}.

On the other hand, to extract the large $z$ behavior, we analyze~\eref{eq:Q1} directly. We note that the function $\mathrm{sech}(2\sqrt{s})$ has simple poles on the negative real axis at $s_n= - (2n+1)^2 \pi^2/16$ with $n=0, 1, 2,\dotsc$. Therefore, large $z$ behavior of $Q(z)$ can be obtained by retaining only the closest branch point to the origin $s_0=-\pi^2/16$ of the integrand in \eref{eq:Q1}, i.e., 
\begin{equation}
Q(z) \sim \frac{\sqrt{\pi}}{2}\, \int_{\gamma- i \infty}^{\gamma+i \infty} \frac{ds}{2\pi i} \, \frac{e^{sz}}{\sqrt{s+\pi^2/16}} = \frac{1}{{2\sqrt{ z}}} \, \exp\left(-\frac{\pi^2\, z}{16} \right).
\label{eq:Q large z}
\end{equation}
We compare this asymptotic exponential tail behavior of $Q(z)$ in~\eref{eq:Q large z} with numerical simulations in \fref{fig:h(V)} and find excellent agreement.

To summarize, the JPDF $P(x_1,x_2,\dotsc,x_N,t)$ for the positions of the $N$ particles in our stochastically heated Brownian gas with a common fluctuating diffusivity $D(t)=B^2(t)$ has the CIID structure given in~\eref{eq:JPDF}, where the PDF $h(V,t)$ of the Brownian functional $V(t)$, given in~\eref{eq:defV}, has the scaling form given in~\eref{eq:hV2}, with the scaling function $Q(z)$ given in~\eref{eq:Q1}. The scaling function $Q(z)$ is given by the exact infinite series in~\eref{Q(z)} and  
has the following limiting behaviors
\begin{equation}
Q(z) \sim \begin{cases}
\displaystyle
\frac{1}{\sqrt{2\pi}\, z^{3/2}}\, e^{-1/(4 z)} &\quad\text{for }~ z\to 0,\\[5mm]
\displaystyle
\frac{1}{2\sqrt{z}}\, e^{-\pi^2 z/16} &\quad\text{for }~ z\to \infty.
\end{cases}
\label{eq:Q-limiting}
\end{equation}

Now, given the CIID structure in~\eref{eq:JPDF}, we next proceed to compute various macroscopic and microscopic observables. In particular, motivated by earlier studies~\cite{BLMS2023, BLMS2024, BKMS2024, SM2024, KMS2025}, we compute the following observables:
\begin{enumerate}
\item Correlation between the particles. 
    \item The average density profile of particles $\rho(x,t)$.
    \item The distribution of the maximum displacement $M_1=\mathrm{max}[x_1(t),x_2(t), \ldots x_N(t)]$, called the extreme value distribution.
    \item The distribution of the $k$-th largest displacement $M_k$, called the order statistics. 
    \item The distribution of the spacing between successive particles $g=M_k-M_{k+1}$, called the gap statistics. 
    \item The distribution of the number of particles in the domain $[-L,L]$, called the full counting statistics (FCS).
    \end{enumerate}

\section{Correlation functions}
\label{s:cor}

The CIID structure in~\eref{eq:JPDF} makes it immediately evident that the random variables $\{x_1, x_2, \dotsc, x_N\}$ are correlated, as the JPDF is not a product of the marginal distributions of each variable. Nevertheless, let us explicitly compute the correlations between the positions of the particles. 
Since the JPDF in~\eref{eq:JPDF|V} is symmetric with respect to any $x_i$ for any given $V(t)$, the mean $\langle x_i\rangle|_V $ vanished even before averaging over $V$, for any $i$. As a consequence the two-particle correlation function $\langle x_i x_j \rangle -\langle x_i \rangle \langle x_j \rangle$ for $i\ne j$ trivially vanishes. Therefore, to capture the two-particle correlations, we need to go to the next order correlation function $C_{i,j}=\langle x_i^2 x_j^2 \rangle - \langle x_i^2 \rangle \langle x_j^2 \rangle$. This correlation function is easy to compute by using the JPDF in  \eref{eq:JPDF}, as
\begin{equation}
C_{i,j}=\langle x_i^2 x_j^2 \rangle - \langle x_i^2 \rangle \langle x_j^2 \rangle = \langle \, V^2(t) \rangle - \langle \, V(t) \, \rangle^2\quad \text{for~} i\ne j.
\label{eq:C_ij}
\end{equation}

Now, using the scaling form~\eref{eq:hV2}, it is quite easy to see that the moments of $h(V,t)$ are given by
\begin{equation}
\langle V^n \rangle = [\xi(t)]^{2n} \, \int_0^\infty  z^n\, Q(z)\, dz = [\xi(t)]^{2n} \,  (-1)^n \frac{d^n}{ds^n} \sqrt{\mathrm{sech}( 2\sqrt{s})} \bigg|_{s=0},
\end{equation}
where $\xi(t)$ is defined in~\eref{eq:hV2} and we have used~\eref{eq:Q1} for $Q(z)$. In particular, we get the first two moments as
\begin{equation}
\langle V\rangle = [\xi(t)]^2 \quad\text{and}\quad \langle V^2 \rangle = \frac{7}{3}[\xi(t)]^4.
\end{equation}
Consequently, from~\eref{eq:C_ij}, we get
\begin{equation}
C_{i,j}= \langle x_i^2 x_j^2 \rangle - \langle x_i^2 \rangle \langle x_j^2 \rangle =\frac{4}{3} [\xi(t)]^4 = \frac{16}{3}\, \Lambda^4\, t^4,
\label{eq:cij}
\end{equation}
where we have used $\xi(t) = \sqrt{2} \, \Lambda \, t$ in the last step.
Therefore, the particles have strong positive correlations between them, which are dynamically generated by the common fluctuating diffusivity. It also follows from~\eref{eq:cij}  that typical $x_i$ scales as $\xi(t)\sim \Lambda t$, indicating that the typical positions of the particles grow ballistically with $t$.

\section{Average density profile of the gas}
\label{s:density}

\begin{figure}
    \centering
    \includegraphics[width=0.6\textwidth]{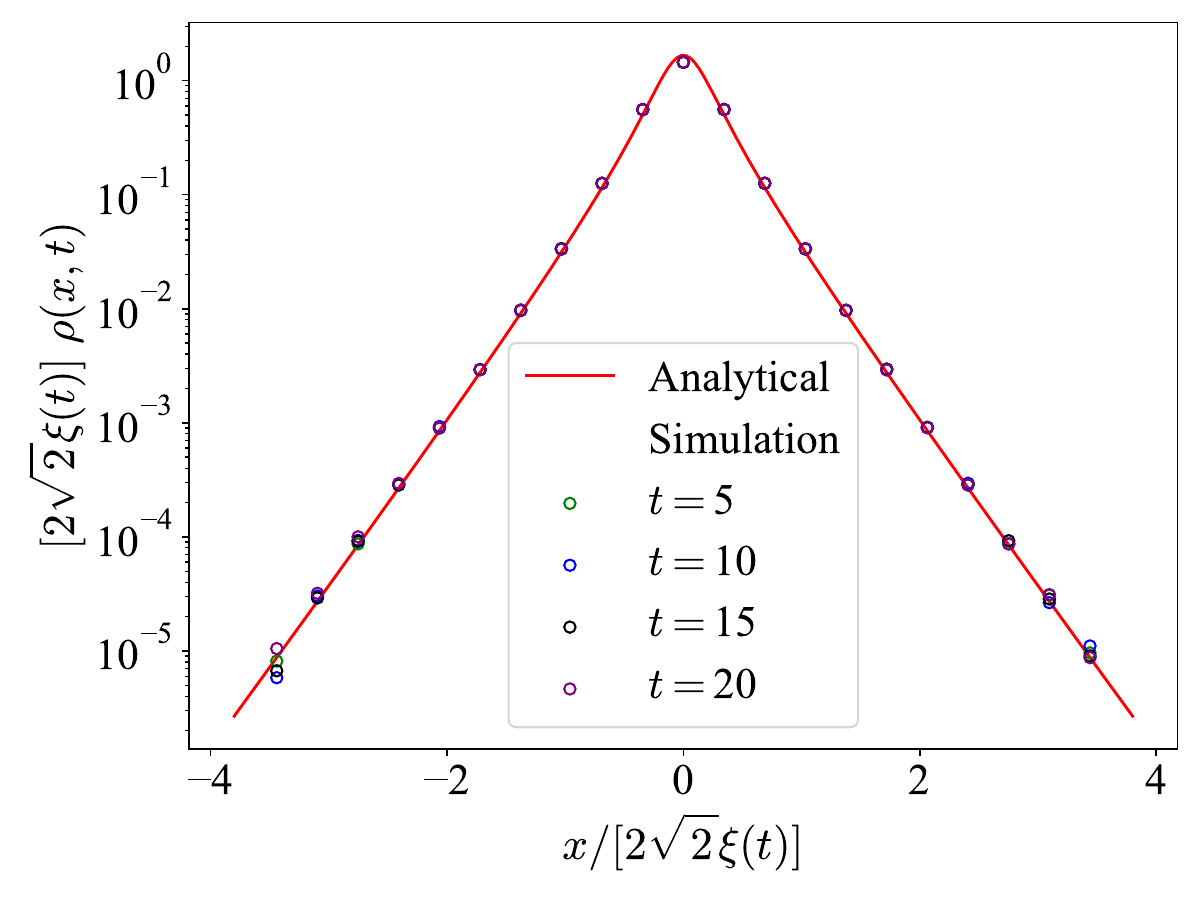} 
   \caption{The scaled collapsed plot of the distribution of the average density profile at various times. The solid curve plots the closed form of the analytical scaling function $f(z) \equiv [2 \sqrt{2} \xi(t)] \, \rho(x,t)$ given in~\eqref{single particle distribution}.
  The parameters used in the simulation are  $\Lambda=1$, $dt = 0.01$, and we average over $10^7$ realizations.  }
    \label{fig:density profile}
\end{figure}

The average density of particles at a position $x$ at time $t$ is defined by
\begin{equation}
    \rho(x,t) = \frac{1}{N} \left\langle  \sum_{n=1}^N  \delta(x_i(t)-x) \right\rangle,
    \label{eq:rho0}
\end{equation}
where $\langle\mathcal{O}\rangle$ represents the average of the observable $\mathcal{O}$, evaluated with respect to the JPDF in~\eref{eq:JPDF} is symmetric with respect to the permutation of the position variables, from~\eref{eq:rho0} we have $\rho(x,t)=\langle \delta(x_1(t)-x)\rangle$. Therefore, the average density profile is, in fact, the marginal PDF $\int_{-\infty}^\infty\dotsm \int_{-\infty}^\infty P(x, x_2, \dotsc, x_N)\, dx_2 \dotsm dx_N$,
\begin{equation}
\rho(x,t) = \int_0^\infty dV\, h(V,t) \, \frac{1}{\sqrt{2\pi V}}\, \exp\left(-\frac{x^2}{2V}\right),
\label{eq:rho1}
\end{equation}
or equivalently, the PDF of the position of a particle for the $N=1$ case. As we mentioned in the introduction, the position distribution of a single Brownian particle with a fluctuating diffusivity $D(t)=B^2(t)$ has been studied earlier~\cite{SGMSS2020, SBS2021, SBS2022}, where it was found that the PDF of the position admits the scaling form~\footnote{The numerical factor $2\sqrt{2}$ in the scaling form is not important and merely included to have a familiar form of the scaling function  based on prior knowledge.}
\begin{equation}
    \rho(x,t) = \frac{1}{2 \sqrt{2} \, \xi(t)} f\left(\frac{x}{2\sqrt{2} \, \, \xi(t)}\right) \, ,
    \label{eq:rho2}
\end{equation}
with the scaling function given by~\cite{SBS2021, SBS2022} (also see~\cite{SGMSS2020} for an alternative expression), 
\begin{gather}
    f(y) = \frac{1}{\sqrt{2 \pi^3}} \, \Gamma\left(\frac{1}{4} + iy\right) \,\Gamma\left(\frac{1}{4}- iy\right)  .
    \label{single particle distribution}
\end{gather} 
In figure \ref{fig:density profile}, we plot the distribution of the scaled density profile obtained from numerical simulation and compare it with the scaling function $f(y)$ in~\eref{single particle distribution} and find excellent agreement. 
The scaling function $f(y)$ in~\eref{single particle distribution} has the following limiting behaviors~\cite{SGMSS2020,SBS2021, SBS2022} 
\begin{equation}
    f(y) \sim  \begin{cases} 
    \displaystyle
    \frac{[\Gamma(1/4)]^2}{\sqrt{2\pi^3}}  \bigl[1-a y^2 +O(y^4) \bigr] &\quad\text{as }~ y\to 0 \\[5mm]
    \displaystyle
    \sqrt{\frac{2}{\pi \, |y|}} \,\, e^{-\pi |y|} & \quad\text{as }~ |y|\to\infty.
    \end{cases}
    \label{asymptotic expansion density scaling function}
\end{equation}
where $\Gamma(1/4)]^2/\sqrt{2\pi^3}=1.66925\dotsc$ and the constant $a= 17.1973\dotsc$

It is worth remarking that,
indeed, using the scaling form of $h(V,t)$ from~\eref{eq:hV2} in~\eref{eq:rho1} suggests the scaling form in~\eref{eq:rho2} with the scaling function given by
\begin{equation}
f(y) = \frac{2}{\sqrt{\pi}}\, \int_0^\infty\, dz\,Q(z)\, \frac{1}{\sqrt{z}}\, e^{-4 y^2/z}.
\label{eq:fy2}
\end{equation}
Now, using the expression of $Q(z)$ from~\eref{Q(z)} in~\eref{eq:fy2} and integrating over $z$, we get an alternative expression for the scaling function $f(y)$ as
\begin{equation}
    f(y) = \frac{4 \sqrt{2}}{\pi} \, \sum_{n=0}^\infty \binom{- \frac{1}{2}}{n} \, \frac{4n+1}{(4n+1)^2+16 y^2} ~.
    \label{expansion for density profile}
\end{equation}
We note that while the small $y$ behavior of $f(y)$ in~\eref{asymptotic expansion density scaling function} can be directly obtained from~\eref{expansion for density profile} by expanding the summand around $y=0$ and performing the summation separately for each order in $y$, the asymptotic tail behavior for large $y$ does not follow from~\eref{expansion for density profile} directly. 
The equivalence between~\eref{single particle distribution} and~\eref{expansion for density profile}  can be easily established by taking a Fourier transform of~\eref{expansion for density profile} and then summing over $n$, which gives
\begin{equation}
\tilde{f}(k) = \frac{1}{\sqrt{2\pi}} \sqrt{\mathrm{sech}(k/2)}.
\label{eq:fFT}
\end{equation}
Now,~\eref{eq:fFT} is nothing but the Fourier transform of the PDF in~\eref{single particle distribution}~\cite{SBS2021}, i.e., 
\begin{equation}
\frac{1}{\pi}\int_0^\infty \cos(k y) \sqrt{\mathrm{sech}(k/2)}\, dk =  \frac{1}{\sqrt{2 \pi^3}} \, \Gamma\left(\frac{1}{4} + iy\right) \,\Gamma\left(\frac{1}{4}- iy\right) .
\end{equation}

\section{Order Statistics}
\label{s:order}

Let $M(t) = \{M_1,M_2,M_3, \ldots, M_N\}$, be an ordered set that describes, at a particular time $t$, the displacements of the particles in a descending order, i.e., $M_1>M_2 > M_3> \dotsb >M_N$, such that $M_1$ denotes the maximum position, $M_2$ denotes the second maximum and so on.   These $M_k$'s are random variables, and the distribution of these is called order statistics. The case of $k=1$ and $k=N$ corresponds to \emph{extreme value statistics} (EVS). 
Before going to the EVS, we look at the order statistics in the bulk, in the limit $N\to\infty$ and  $k\to\infty$, keeping $\alpha = k/N 
\in (0,1)$. 
The probability that the $k$-th element assumes the value $w$ for a particular realization of $\{D(\tau)\}$, i.e., for a given $V$,  is given by~\cite{Sabhapandit_2008,  majumdar2020extreme, BLMS2023, majumdar2024statistics} ,
\begin{equation}
    \mathrm{Prob.}(M_k = w|V) = \frac{N!}{(k-1)! \, (N-k)!} \, [1-\psi(w|V)]^{N-k} \, p(w|V) \, [\psi(w|V)]^{k-1} \ ,
\end{equation}
where 
\begin{equation}
p(x|V)=\frac{1}{\sqrt{2\pi V}}\, \exp\left(-\frac{x^2}{2V}\right)\quad [\text{see}~\eref{eq:JPDF}]
\label{eq:pxv}
\end{equation}
is the conditional PDF of the position of a single particle for a given $V$ (the conditional positions of the particles are independent for a given $V$), and  
\begin{equation}
\psi(w|V)= \int_{w}^{\infty} dx' p(x'|V) = \frac{1}{2}\, \mathrm{erfc}\left(\frac{w}{\sqrt{2 V}}\right)
\label{eq:psi}
\end{equation}
is the conditional cumulative distribution. 
 Now, using Stirling's approximation for the bulk, one arrives at 
\begin{equation}
    \mathrm{Prob.}(M_k = w|V) \approx \sqrt{\frac{N \alpha}{ 2 \pi (1-\alpha)}} \, \frac{p(w|V)} {\psi(w|V)} \,  e^{-N \phi(w)} \, ,
\end{equation}
where  $\phi(w)$ given by,
\begin{equation}
    \phi(w) = \alpha \ln\left[\frac{\alpha}{\psi(w|V)}\right] + 
    (1-\alpha) \ln\left[\frac{1-\alpha}{1-\psi(w|V)}\right] .
    \label{large deviation function}
\end{equation}
The cumulative distribution $\psi(w|V)$ in~\eref{eq:psi} is a monotonically decreasing function of $w$, with $\psi(w|V)\to 1$ as $w\to -\infty$ and $\psi(w|V)\to 0$ as $w\to \infty$. Consequently, for any $\alpha\in(0,1)$, the large deviation function $\phi(w|V)$ is a convex function that diverges at both ends $w\to\pm\infty$ and has a unique minimum at a particular value $w=q(\alpha, V)$. The location of the minimum can be obtained by setting $\phi'(q)=0$, which, in turn, gives $\psi(q|V)=\alpha$. Therefore, using the explicit expression from~\eref{eq:psi}, we have
\begin{equation}
    q(\alpha, V) = \sqrt{2 \, V} \, \mathrm{erfc}^{-1} (2 \alpha) \ ,
    \label{qdefn}
\end{equation}
where $\mathrm{erfc}^{-1}$ is the inverse of the complementary error function. As $\alpha$ varies continuously from $0$ to $1$, the location of the minimum $q(\alpha, V)$ decreases monotonically from $\infty$~to~$-\infty$, crossing zero at $\alpha=1/2$.

It is evident from~\eref{large deviation function} that  $\phi(q)=0$. Moreover, from~\eref{large deviation function}, using~\eref{eq:psi} and $\psi(q|V)=\alpha$, it is easy to find that $
\phi''(q)=[p(q|V)]^2/[\alpha(1-\alpha)]$. Therefore, 
expanding $\phi(w)$ around $q$ up to the second order one finds that 
\begin{equation}
    \mathrm{Prob.}(M_k =w|V) \approx \sqrt{\frac{N [p(q|V)]^2 }{2 \pi \alpha (1-\alpha)}} \,  \exp{\bigg[ -N \frac{ [p(q|V)]^2}{2 \alpha (1-\alpha)}\,  (w-q)^2\bigg]} \, , \label{order statistics, expansion of phi}
\end{equation}
where $p(q|V)$ is defined in~\eref{eq:pxv}.
In the limit $N  \rightarrow \infty$, the variance of this Gaussian distribution tends to zero, and it converges to a Dirac delta function, 
\begin{equation}
   \lim_{N\to\infty} \mathrm{Prob.}(M_k = w|V) = \delta[w- q(\alpha,V)] \, .
\end{equation}

\begin{figure}
    \centering    
    \includegraphics[width=0.6\textwidth]{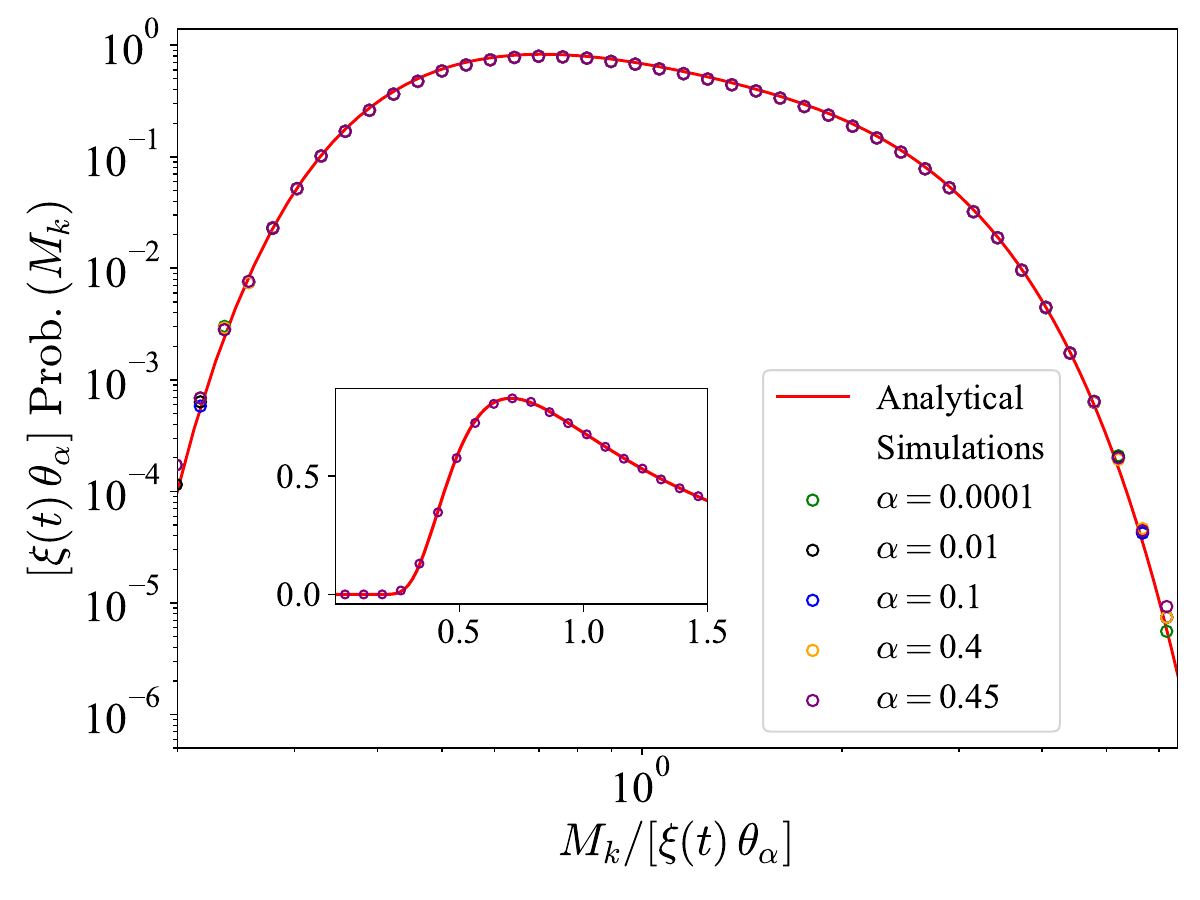} 
    \caption{
       The PDF of the scaled variable $M_k/[\xi(t) \theta_\alpha]$ for various $\alpha = k/N$. The points represent the simulation data for parameters   $\Lambda =1$, $dt = 0.01$, $t=4$, $N= 10^6$, and averaged over $  10^6$ realizations. The solid curve is the analytical scaling function $Q(z)$ described in equation \eqref{order_statistics equation}.
       The inset shows the essential singularity of the distribution near zero. 
     }
    \label{fig:order stat scaled_plot}
\end{figure}

Therefore, averaging over the distribution $h(V,t)$ given in~\eref{eq:hV2}, we obtain the unconditional order statistics as
\begin{equation}
    \mathrm{Prob.}(M_k = w) = [\xi(t)]^{-2}\,  \int_0^\infty dV \, Q\left(V/ [\xi(t)]^2\right) \,\delta[w -q(\alpha,V)] \,  .
    \label{Order statistics integral h(V)}
\end{equation}
 Since $q(\alpha,V)$ in~\eref{qdefn} is positive for $\alpha\in(0,0.5)$ and any $V$, the distribution in~\eref{Order statistics integral h(V)} is supported over $w\in (0,\infty)$ for $\alpha\in(0,0.5)$.  Similarly, the distribution is supported over $w\in (-\infty, 0)$ for $\alpha\in(0.5,1)$. Using the explicit expression of $q(\alpha,V)$ from~\eref{qdefn}, the integral in~\eref{Order statistics integral h(V)} can be performed explicitly, giving 
\begin{equation}
    \mathrm{Prob.}(M_k = w) =  \frac{|w|}{ [\theta_\alpha\,\xi(t)]^2}  \  Q\bigg (\frac{w^2}{2 [\theta_\alpha\,\xi(t)]^2} \bigg) \ ,\quad\text{with}\quad  \theta_\alpha = \mathrm{erfc}^{-1}(2 \,\alpha) ,
    \label{order_statistics equation} 
\end{equation}
where $Q(z)$ is defined in~\eref{Q(z)}. 
We compare~\eref{order_statistics equation} with numerical simulations in~\fref{fig:order stat scaled_plot}
and find excellent agreement. We remark that the the limiting behavior of $Q(z)$ in~\eref{eq:Q-limiting} translates to the following limiting behaviors of $\mathrm{Prob.}(M_k = w)$ in~\eref{order_statistics equation},
\begin{equation}
 \mathrm{Prob.}(M_k = w)  \sim \begin{cases}
 \displaystyle
 \frac{2\, \xi(t) |\theta_\alpha|}{\sqrt{\pi}\, w^2}\, \exp\left(-\frac{[\xi(t)]^2\theta_\alpha^2}{2 w^2}\right) &\quad\text{as }~ |w|\to 0,\\[5mm]
  \displaystyle
  \frac{1}{\sqrt{2}\, \xi(t)\, |\theta_\alpha|}\, \exp\left(-\frac{\pi^2w^2}{32[\xi(t)]^2\theta_\alpha^2}\right) &\quad\text{as }~ |w|\to \infty.
 \end{cases}
\end{equation}

\subsection{Position fluctuation of  the central particles ($\alpha = 0.5$):} 
\label{s:central}

\begin{figure}
    \centering    
    \includegraphics[width=0.6\textwidth]{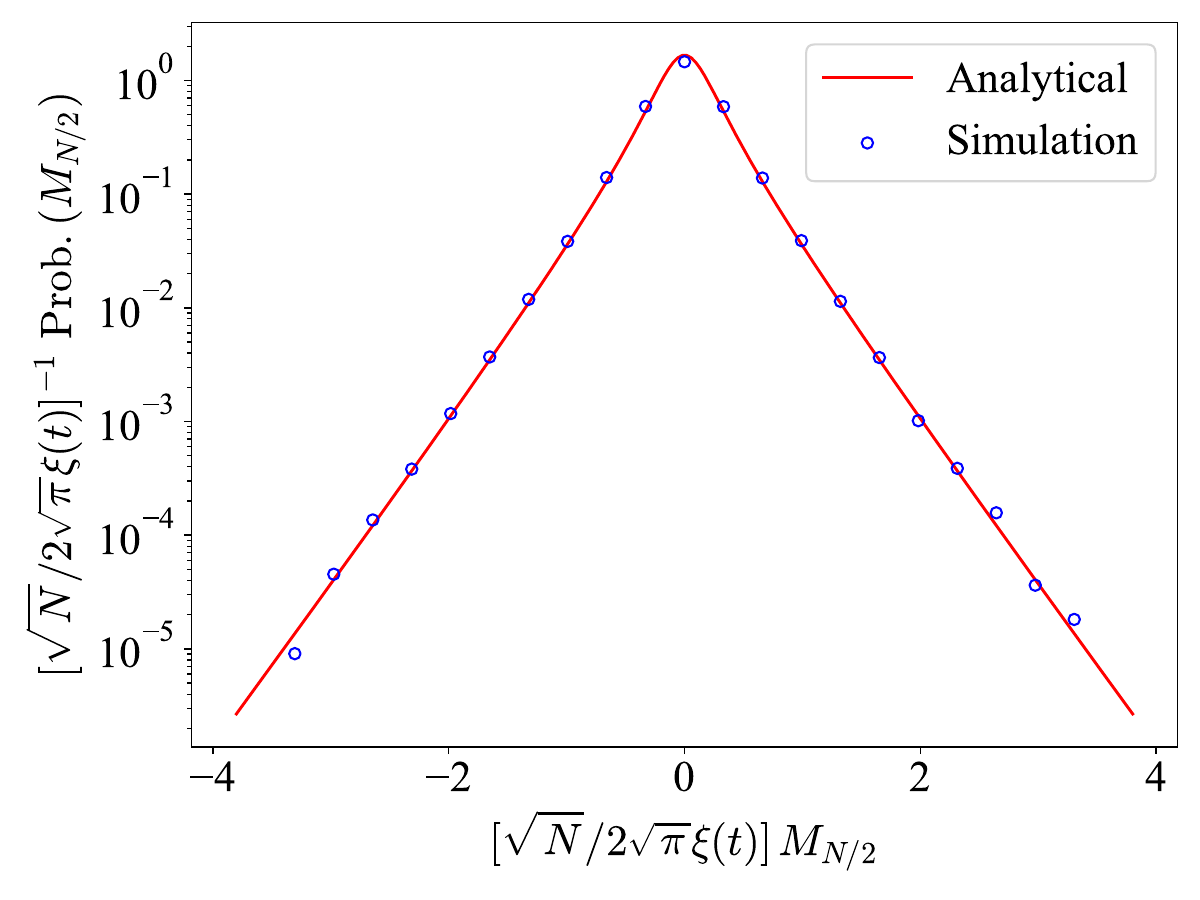} 
    \caption{The PDF of the central particle with the appropriate scaling. The points represent the simulation data for the parameters 
            $\Lambda =1$, $dt = 0.01$, $t=4$, $N= 10^6$, and averaged over $ 10^6$ realizations. The solid line plots the closed form of the scaling function $f(z) \equiv  \,[\sqrt{N}/ (\, 2 \sqrt{\pi} \xi(t) \, )]^{-1} \,  \mathrm{Prob}.(M_{1/2}) $ given in equation \eqref{single particle distribution}.}
    \label{fig:central particle_plot}
\end{figure}

The PDF of the $k$-th position given in~\eref{order_statistics equation} is expected to be valid for particles in the bulk. For  the special case $\alpha=1/2$, i.e., for the central particle,  from~\eref{qdefn} 
$q(1/2,V)=0$.  Therefore, from~\eref{Order statistics integral h(V)}, the PDF $\mathrm{Prob}.(M_{N/2}=w)\to \delta(w)$ to the leading order.
To obtain the finite $N$ correction for large $N$, 
we use~\eqref{order statistics, expansion of phi} with $\alpha=1/2$ and average over $V$ using $h(V,t)$ from~\eref{eq:hV2},
\begin{equation}
    \mathrm{Prob.}(M_{N/2}=w) = \frac{\sqrt{N} }{\pi  } \int_0 ^\infty dV \, \frac{Q( V/[\xi(t)]^2) }{[\xi(t)]^2\,\sqrt{V}} \, \exp{\bigg[ - \frac{N w^2}{\pi V }\bigg]} \ ,
\end{equation}
where we have used $p(0|V) = 1/\sqrt{2 \pi V}$. Making a change of variable $z= V /[\xi(t)]^{2}$ gives, 
\begin{equation}
    \mathrm{Prob.}(M_{N/2}=w) =  \frac{\sqrt{N}} {\pi \,\xi(t)} \, \int_0 ^\infty dz \, \frac{Q(z)}{\sqrt{z}}\,  e^{- 4y^2/z} 
   \quad\text{with}\quad y=\frac{\sqrt{N} \,w}{2\sqrt{\pi}\, \xi(t)} ~.
    \label{Central distribution diffusing diffusivity form} 
\end{equation}
Now we note that the integral in~\eref{Central distribution diffusing diffusivity form} is the same as in~\eref{eq:fy2}. Therefore, after evaluating the integral, we get
\begin{gather}
    \mathrm{Prob.}(M_{N/2}=w) = \frac{\sqrt{N}} {2 \sqrt{\pi}\, \xi(t)} \  f\left(\frac{\sqrt{N} \,w} {2 \sqrt{\pi} \,\xi(t)}\right) \, ,
    \label{Order statistics finite size central}
\end{gather}
where the scaling function $f(z)$ is given in~\eref{single particle distribution}. 
The asymptotic form of the scaling function $f(y)$ in~\eref{asymptotic expansion density scaling function} gives an exponential tail for the PDF~\eref{Order statistics finite size central} of the position of the central particle,
 \begin{equation}
    \mathrm{Prob}. \, (M_{N/2} =w) = 
    \left( \frac{\sqrt{\pi N}}{\xi(t)}\right)^{\frac{1}{2}} \, \frac{1}{\pi \sqrt{|w|}} \, \exp\left( - \frac{\sqrt{\pi N}}{2\ \xi(t)} \, |w| \right) \, .
    \label{eq:middle-scaling}
\end{equation}
In \fref{fig:central particle_plot}, we compare~\eref{Order statistics finite size central} with numerical simulations and find excellent agreement. 
The typical width of the distribution  $\sqrt{\langle w^2 \rangle }\sim \xi(t)/\sqrt{N}$ tends to zero in the limit $N\to \infty$ for any finite $t$, forcing the distribution in~\eref{Order statistics finite size central} to a Dirac delta function. 

It is interesting to note that the scaling function $f(y)$ in~\eref{Order statistics finite size central} describing the PDF of the position of the middle particle is exactly the same as the scaling function that appears in the average density profile in~\eref{eq:rho2}.
Mathematically, this universality arises because of the equivalence of the integrals in~\eref{eq:fy2} and~\eref{Central distribution diffusing diffusivity form}, for any $Q(z)$. The function $Q(z)$, of course, depends on specific models, resulting in different forms of $f(y)$ for different models. For example, $Q(z)=re^{-rz}$ for a Brownian gas correlated by resetting~\cite{BLMS2023},  whereas it has a different expression for a Brownian gas correlated via a switching harmonic trap~\cite{BKMS2024}. Nevertheless, the equivalence between the position distribution of the central particle and the average density profile remains valid for a large class of models where the JPDF has the particular CIID structure given in~\eref{eq:JPDF}.

\subsection{Distribution of extreme values:} 
\label{Section:EVS}

Let us now look into the distribution of the position of the rightmost particle $M_1=\max(\{x_1, x_2, \dotsc,x_N\})$ separately. It is more convenient to consider the cumulative distribution of $M_1$. The probability that $M_1<w$ is mathematically equal to the probability that all the positions  $\{x_1, x_2, \dotsc,x_N\}$ are less than the value $w$. The latter, for a given realization of the diffusion coefficient, i.e., for a given  $V$, can be expressed as~\cite{majumdar2020extreme, sabhapandit2019extremes, Sabhapandit_2008, majumdar2024statistics}  
\begin{equation}
    \mathrm{Prob}. ( M_1<w|V) = \left[1-\psi(w)\right]^N, 
\end{equation}
where $\psi(w)$ is given in~\eref{eq:psi}. Since $\psi(w)\to 0$ for $w\to\infty$,  taking the limit $w\to\infty$ and $N\to \infty$, keeping $N\psi(w) \sim O(1)$, we get
\begin{equation}
    \mathrm{Prob} .(M_1<w|V)  \to  \exp\left[-N \psi(w) \right]  . 
    \label{EVS2}
\end{equation}

For large $N$, we expect the typical location of the maximum $\sqrt{2V}a_N$ to be such that the argument of the exponential $N\psi(\sqrt{2V}a_N)$ in~\eref{EVS2} to be $O(1)$, where $\sqrt{2V}$ is a trivial scaling due to the form of $\psi(w|V)$ in~\eref{eq:psi}. From~\eref{eq:psi}, this means $a_N\sim \sqrt{\ln N}$. Therefore, for large $N$, the maximum value scales as $M_1=\sqrt{2V}\, (a_N +b_N z)$, where $\sqrt{2V}\, b_N$ denotes the strength of the fluctuations and $z$ is a random variable describing the PDF of the fluctuations around $\sqrt{2V}\, a_N$. Using $\psi(w)$ from~\eref{eq:psi} and taking the limit $N\to\infty$ we arrive at the Gumbel distribution~\cite{majumdar2020extreme, sabhapandit2019extremes, Sabhapandit_2008, majumdar2024statistics} 
\begin{equation}
\lim_{N\to\infty}  \mathrm{Prob}. ( M_1<w= \sqrt{2V}(a_N +b_N z)|V)  \rightarrow
e^{-e^{-z}}.
\label{eq:gumbel}
 \end{equation}
In order for the right hand side of~\eref{eq:gumbel} to be independent of $N$, the parameters $a_N$ and $b_N$,  must satisfy the following relations
\begin{equation}
  a_N   = \mathrm{erfc}^{-1}(2/N) \sim \sqrt{\ln N }\quad\text{and}\quad 
    b_N =\frac{\sqrt{\pi}}{N}\, e^{a_N^2} \sim \frac{1}{2\sqrt{\ln N} }.
  \label{eq:ab}
\end{equation}
The strength of the fluctuation $b_N\to 0$ as $N\to \infty$, and therefore, to the leading order, one can neglect the fluctuations and get
\begin{equation}
\mathrm{Prob.}~(M_1=w|V) = \delta(w-\sqrt{2V}a_N).
\label{eq:evs3}
\end{equation}

The unconditional distribution for the extreme value $M_1$ is given by
 \begin{equation}
     \mathrm{Prob.}(M_1 = w)= \int_{0}^{\infty} dV \, h(V,t) \,  \mathrm{Prob}.(M_1=w|V)  .
     \label{eq:evs:5}
 \end{equation}
Therefore, using~\eref{eq:hV2} and~\eref{eq:evs3} in~\eref{eq:evs:5}, we get
 \begin{equation}
     \mathrm{Prob.}(M_1 = w) = \int_{0}^{\infty} \frac{dV}{[\xi(t)]^2} \, Q\left( \frac{V}{\xi(t)} \right)   \delta(w- \sqrt{2 V } a_N) 
     = \frac{w}{[a_N\,\xi(t)]^2} \, Q\bigg(\frac{w^2}{2 [a_N \,\xi(t)]^2} \bigg) \ .\label{evdist}
\end{equation}
Therefore, the PDF of the maximum $M_1$ in~\eref{evdist} is a special case of the order statistics in~\eref{order_statistics equation} with $\alpha=1/N$, as 
$a_N$ defined in~\eref{eq:ab} is the same as $\theta_{1/N}$ defined in~\eref{order_statistics equation}.

\section{Gap statistics}
\label{s:gap}

\begin{figure}
    \centering
    \includegraphics[width=0.6\textwidth]{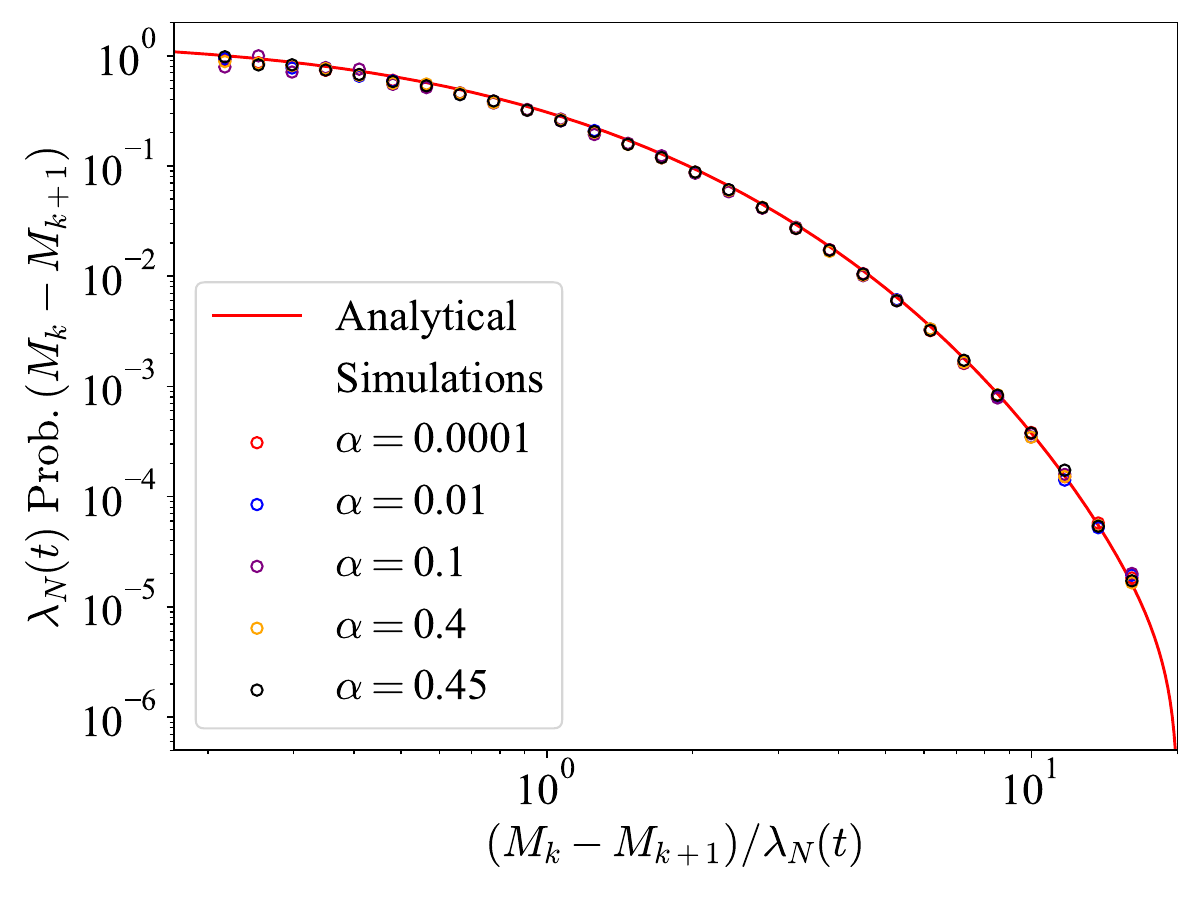} 
    \caption{Plot of the distribution of the scaled spacing variable $  (M_k - M_{k+1})/\lambda_N(t)$.  The points are from numerical simulations while the solid line plots the scaling function $F(z)$ in \eref{eq:Hx}. The  parameters used in the simulation are $\Lambda =1$, $dt=0.01$, $t =4$,  $N = 10^6$, and  averaged over $10^6$ realizations.}
    \label{fig:Scaled_spacing_distribution}
\end{figure}

While the order statistics describes the position fluctuations of individual ordered particles, the gap statistics describes the statistics of separations between successive ordered particles. Let $d_k(V)=M_k(V) - M_{k+1}(V)$ denotes the gap between the $k$-th and $(k+1)$-th ordered particles, for a given realization of $V$. Then the unconditional gap statistics is given by
\begin{equation}
    \mathrm{Prob.}(d_k=g) = \int_{0}^{\infty} dV\, h(V,t)\,
    \mathrm{Prob.} \bigl(M_k(V)-M_{k+1}(V)=g\bigr).
    \label{dk}
\end{equation}
The conditional PDF $\mathrm{Prob.} \bigl(M_k(V)-M_{k+1}(V)=g\bigr)$ can be obtained from the joint distribution of $M_k$ and $M_{k+1}$ for a given $V$~\cite{BLMS2023},
\begin{align}
\mathrm{Prob.}[M_k(V)=x, M_{k+1}(V)=y] = \frac{N!}{(k-1)! (N-k-1)!}\, p(x|V) p(y|V)\cr
\times \left[\psi(x|V)\right]^{k-1}
    \left[1-\psi(y|V)\right]^{N-k-1} \theta(x-y).
    \label{eq:JointMM}
    \end{align}
    The distribution of the gap $d_k(V)= M_k(V) - M_{k+1}(V)$ for a given $V$ is then given by 
   \begin{equation}
    \mathrm{Prob.}(M_k(V)- M_{k+1}(V)=g) =\int_{-\infty}^\infty  \mathrm{Prob.} \bigl[M_k(V) = y+g, M_{k+1}(V) = y\bigr]\, dy. 
    \label{eq:gap1}
   \end{equation} 
 Starting with the joint distribution in~\eref{eq:JointMM} and going via~\eref{eq:gap1}, in the large $N$ limit, one finds that~\cite{BLMS2023}    
\begin{equation}
\mathrm{Prob.}\left(M_k(V) \, -M_{k+1}(V) \, = \, g  |V \right) \approx N \, p(q | V) \, e^{-N \, p(q | V )\, g} \ , 
\label{general spacing distribution}
\end{equation}
where $q \equiv q(\alpha,V) = \sqrt{2V} \,  \mathrm{erfc}^{-1}(2 \alpha)$ defined in equation~\eqref{qdefn}. 
Finally, averaging equation~\eqref{general spacing distribution} over $h(V,t)$, as  in~\eref{dk}, and using the  
scaling form in~\eref{eq:hV2} and $p(x|V)$ in~\eref{eq:pxv}, we get
\begin{equation}
    \mathrm{Prob.}(d_k= g) = \frac{1}{\lambda_N(t)} \, F\bigg(\frac{g}{\lambda_N(t)}\bigg) \, ,
\end{equation}
where $\lambda_N (t) =  \sqrt{2 \pi} \, \xi(t) \ e^{\theta^2_\alpha}/N$ with $\theta_\alpha=\mathrm{erfc}^{-1}(2 \alpha)$ and the scaling function $F(z)$ is given by
\begin{equation}
    F(z) = \int_{0}^\infty du \, Q(u)\,  \frac{e^{-z/\sqrt{u}}}{\sqrt{u}} \ .\label{scaling function gap statistics 1}
\end{equation}
Using the expression~\eref{Q(z)} for $Q(z)$,  the integral in~\eref{scaling function gap statistics 1} can be performed explicitly, which  gives,
\begin{equation}
     F(z) = \frac{2 \sqrt{2}}{\sqrt{\pi}} \, \sum_{n=0}^{\infty} \binom{-\frac{1}{2}}{n} \frac{1}{c_n} \, \left[1 - \frac{ \sqrt{\pi} \, z \ e^{z^2/c_n^2}}{c_n}  \ \mathrm{erfc}\left(\frac{z}{c_n}\right) \right] \quad \text{with}\quad
    c_n= 4n+1.
      \label{scaling function gap statistics 2}
\end{equation}
The two infinite sums are individually convergent, and the first can be  explicitly evaluated,
\begin{equation}
    F(z) = \frac{[\Gamma(1/4)]^2}{2 \sqrt{2} \, \pi} - 2 \sqrt{2} \, \sum_{n=0}^\infty \, \binom{-
    \frac{1}{2}}{n} \, \frac{z \ e^{z^2/c_n^2}}{c_n^2} \, \mathrm{erfc}{\bigg(\frac{z}{c_n} \bigg)} \ .
    \label{eq:Hx}
\end{equation}
The scaling function $F(z)$ has support over the positive real line. It describes the gap statistics in the bulk as well as the edge. We compare this analytical result with numerical simulations in 
\fref{fig:Scaled_spacing_distribution} and find excellent agreement. The function $F(z)$ approaches a constant as $z\to 0$, 
\begin{equation}
    F(0) = \frac{[\Gamma(1/4)]^2}{2 \sqrt{2} \,\pi} \\
     = 1.47933\dotsc
\end{equation}
and has a linear decay for small $z$,
\begin{equation}
F(z)= F(0) - F_1\, z +O(z^2)\quad\text{with }~F_1=2.78143\dotsc.
\end{equation}
This behavior indicates that there is no effective repulsion between the particles.

Let us now find the tail behavior of $F(z)$. Because of the presence of the essential singularity $e^{-z/\sqrt{u}}$ in the integral in~\eref{scaling function gap statistics 1}, the integrand contributes substantially only for $u \gtrsim  z^2$. Therefore, for large $z$,  
the dominant contribution to the integral in~\eref{scaling function gap statistics 1} comes from the tail behavior of $Q(u)$. Using the large $u$ asymptotic form~\eref{eq:Q large z} for $Q(u)$ in~\eref{scaling function gap statistics 1}, we get
\begin{equation}
    F(z) = \frac{1}{2} \int_0 ^\infty dz \frac{1}{u} \, e^{-u \pi^2/16} \, e^{-z/\sqrt{u}}
    \sim
   \frac{2\pi^{1/6}}{\sqrt{3}\, z^{1/3}}\, e^{-\frac{3}{4} (\pi z)^{2/3}}.
 \label{Spacing distribution tail}
\end{equation}
This stretched exponential tail behavior $F(z)$ in~\eref{Spacing distribution tail} for large $z$ is similar to that observed for the resetting Brownian motion~\cite{BLMS2023}, originating from the exponential tail of $Q(u)$ in both cases.

\section{Full Counting Statistics}
\label{s:FCS}

Finally, we consider the full counting statistics (FCS), which describes the probability distribution of the number of particles in a given region in space. For simplicity, here, we consider the region $[-L,L]$. For a given $V$, since the conditional JPDF of the positions in~\eref{eq:JPDF|V} have the product form,  
the probability of finding $N_L$ particles in between $[-L,L]$, for a given $V$, is simply the binomial distribution,
\begin{equation}
    \mathrm{Prob.}(N_L,N|V) = \binom{N}{N_L} \left[\int_{-L}^L {dx \, p(x|V)}\right]^{N_L} \, \,  \left[1-\int_{-L}^L {dx \, p(x|V)}\right]^{N-N_L} \label{fulll counting binomial}
\end{equation}
In the asymptotic limit, we expect $N_L$ to be large for any non-zero $L$. Therefore, setting $N_L = \kappa N$ with $\kappa\in [0,1]$,  we note that for large $N$, the binomial distribution converges to the Gaussian distribution with mean $\langle N_L \rangle$ and variance $\langle{N^2_L} \rangle_c$ given by,
\begin{gather}
    \langle N_L \rangle = N  \int_{-L}^L dx \,  p(x|V) ,\\
    \langle N^2_L \rangle_c = N \left[\int_{-L}^L dx \, p(x|V)\right] \, \bigg[ 1- \int_{-L}^L dx \, p(x|V)\bigg] .
\end{gather}
Consequently, the relative fluctuations scale as $\sqrt{ \langle N^2_L \rangle_c}/\langle N_L \rangle \sim N^{-1/2}$ and in the limit $N\to \infty$, the distribution for the fraction of particles $\kappa$ in  $[-L,L]$ converges to a Dirac delta function about
\begin{equation*}
        \int_{-L}^L {dx \, p(x|V)} = \mathrm{erf}\left(\frac{L}{\sqrt{2 \, V}}\right) \ .
\end{equation*}
Hence, the probability distribution of $N_L$, which can be found by averaging  $\mathrm{Prob.} (N_L,N|V)$ with respect to $V$ using~\eref{eq:hV2}, has the scaling form
\begin{equation}
 \mathrm{Prob.}(N_L,N, t) = \frac{1}{N}\, H\left(\frac{N_L}{N}, t\right), 
\end{equation}
where the scaling function $H(\kappa, t)$ describing the PDF the fraction of particles $\kappa$ in  $[-L,L]$ at time $t$ is given by
\begin{equation}
   H(\kappa, t) = [\xi(t)]^{-2} \int_{0}^{\infty} dV \, Q(V/ [\xi(t)]^{2}) \, \delta \left[\kappa-\mathrm{erf} \left(\frac{L}{\sqrt{2 V}} \right) \right] \ .
\end{equation}
Performing the above integral leads to,
\begin{equation}
H(\kappa, t)=  \frac{\sqrt{\pi}}{2 [\ell(t)]^2} \,  
     \frac{ \exp[\mathrm{erf}^{-1}(\kappa)^2]}{[\mathrm{erf}^{-1}(\kappa)]^3} \, \, 
      Q\left(\frac{1}{\bigl[\sqrt{2}\,\ell(t)\,\mathrm{erf}^{-1}(\kappa)\bigr]^2} \right)   ,
     ~\text{with}~~ \ell(t)  = \frac{\xi(t)}{L} \propto t,
     \label{FCS} 
\end{equation}
where $\xi(t) =\sqrt{2}\, \Lambda t$, is defined in~\eref{eq:hV2}.
The distribution in~\eref{FCS} is supported over the entire range $[0,1]$. Using $\mathrm{erf}^{-1}(\kappa) = \sqrt{\pi} \kappa/2 +O(\kappa^2)$ as $\kappa\rightarrow0$ and the asymptotic form of $Q(z)$ from~\eref{eq:Q large z} in~\eref{FCS} one finds,
\begin{equation}
  H(\kappa, t) = \sqrt{\frac{2}{\pi}} \, \frac{1}{\ell(t)\,\kappa^2} \, \exp{\left[-\frac{\pi }{8  [\ell(t)]^2\,\kappa^2}\right]} \quad\text{as}\quad \kappa\to 0.
  \label{eq:smallK}
\end{equation}
Thus, the distribution has an essential singularity at the lower support $\kappa=0$. Since all the particles start from the origin, the probability of finding no particles inside the domain $[-L,L]$ is evidently zero at $t=0$. As time progresses, since the gas expands ballistically, more and more particles leave the finite domain $[-L,L]$. This is reflected in the scale at which the essential singularity manifests in~\eref{eq:smallK}, which decreases with time as  $1/\ell(t) \sim 1/t$. 

\begin{figure}
    \centering    \includegraphics[width=\textwidth]{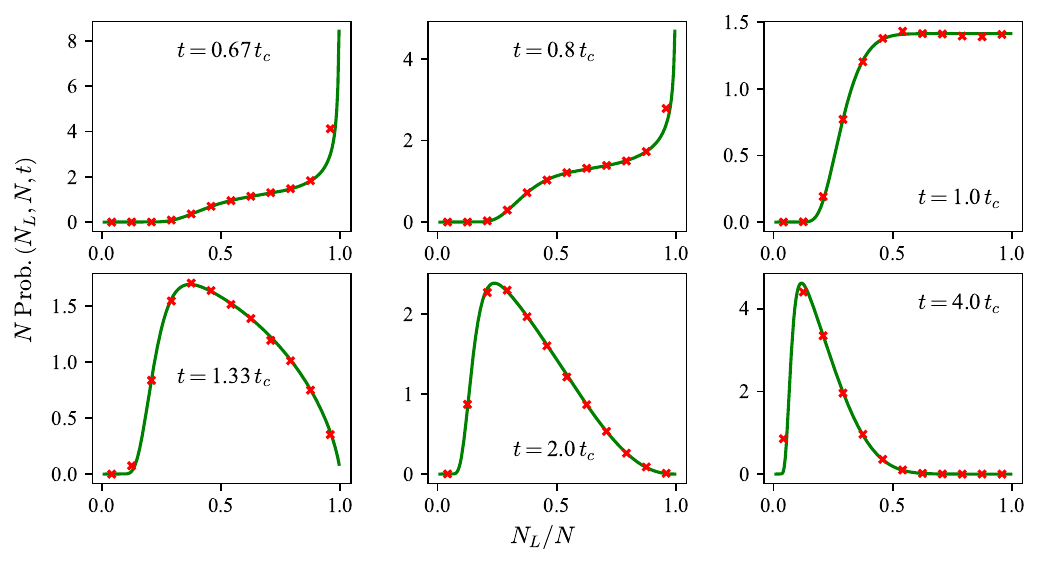}
    \caption{The PDF of the fraction of particles $N_L/N$ in the region $[-L,L]$. The markers are of numerical simulations with parameters   $\Lambda =1$, $N=3 \times 10^5$, averaged over $2 \times 10^5$ realizations. The solid lines plots the function $N \, \mathrm{Prob.}(N_L, N, t) \equiv H(N_L/N \, ,t)$ given in \eqref{FCS}. 
    We see that the curve flattens near $N_L/N =1$ for $t = t_c = L/ \Lambda$. For times less than $t_c$, the probability density approaches zero, while for times greater than $t_c$, there is a power-law divergence at the upper support.} 

    \label{fig:Counting statistics}
\end{figure}

On the other hand, for $\kappa\to 1$ from below, using $\mathrm{erf}^{-1}(\kappa) \sim \sqrt{-\ln[1-\kappa]}$ and the small $z$ behavior of  $Q(z)$ from~\eref{eq:Q-limiting}, we get
\begin{equation}
  H(\kappa, t ) = \ell(t) \, (1-\kappa)^{\beta(t)} \quad\text{with }~ \beta(t) = {\frac{[\ell(t)]^2}{2}-1} = \left(\frac{\Lambda}{L}\right)^2 t^2 -1.
\end{equation}
Thus near the upper support $\kappa=1$, the scaling function 
 $H(\kappa,t)$ has a rather unusual behavior: $H(\kappa,t)\sim (1-\kappa)^{\beta(t)}$ where the exponent $\beta(t)$ changes continuously with time. The scaling function  $H(\kappa,t)$  undergoes an interesting dynamic shape transition near the upper support $\kappa=1$ with time. At early  times $t< t_c=L/\Lambda$,  the exponent $\beta(t)<0$ and consequently, there is a power law divergence of $H(\kappa,t)$ near $\kappa=1$, while for $t>t_c$, $H(\kappa,t)\to 0$ as $\kappa\to 1$. At the transition time $t=t_c$, the function $H(\kappa,t_c)$   becomes flat and converges to a non-zero value as $\kappa\to 1$. 
 This is because, since the gas is initially localized at the origin, all the particles are within $[-L,L]$ at $t=0$. As time progresses, particles leave the region. The time $t_c=L/\Lambda$ is indeed the typical time needed for a particle to hit the boundary $L$ starting at the origin.  
The above results are verified by the simulations presented in \fref{fig:Counting statistics}.

\section{First-passage observables}
\label{sec:FPT}

We now turn to the first-passage properties of the system~\cite{Redner_2001, Bray01062013}. Such properties have previously been investigated for a single Brownian particle with stochastic diffusivity~\cite{jain2016diffusingsurvival, Lanoiselée2018, Grebenkov_2019,target_search, SBS2021, Grebenkov_2021}. 
Here, we move beyond the single-particle case~\cite{Sabhapandit_2007, Krapivsky_2010, mortality, MSS16} and consider a system of 
$N$ particles evolving under a common stochastic diffusivity $D(t)$.

For a given realization of the stochastic diffusivity $\{D(\tau); 0\le \tau \le t\}$, if we make a change of variable from time $t$ to $V$ as,
\begin{equation}
    V = 2 \int_0^t  d\tau \, D(\tau)  \, ,
    \label{eq:V1}
\end{equation}
then, in terms of $V$, the Langevin equation~\eref{eq:LE}   becomes
\begin{equation}
    \frac{dx_j}{dV}  = \beta_j(V) \, ,\quad\text{where}\quad  \beta_j(V) =  \frac{\eta_j(t)}{\sqrt{2 D(t)}}.
    \label{eq:LE_transformed}
\end{equation}
Since the diffusion coefficient $D(t)>0$, the variable $V$ in~\eref{eq:V1} is a monotonically increasing function of $t$, and there is a one-to-one correspondence between the variables $t$ and $V$. In particular, from ~\eref{eq:V1}, one gets 
\begin{equation}
  \delta(V-V') = \frac{\delta(t-t')}{|\frac{dV}{dt}|}= \frac{\delta(t-t')}{2 D(t)}. 
  \label{eq:delV}
\end{equation}

Since the noises $\{\eta_j\} $ are Gaussian, for any given realization $\{D(\tau); 0\le \tau \le t\}$, the noises $\{\beta_j(V)\}$ in~\eref{eq:LE_transformed} are also Gaussian. From \eref{eq:LE_transformed} and \eref{eq:delV}, it immediately follows that 
$\langle \beta_i(V)\rangle =0$ and $\langle \beta_i(V)\beta_j(V') \rangle =\delta_{i,j} \, \delta(V-V')$, for any given realization of $\{D(\tau); 0\le \tau \le t\}$. Therefore, the positions $\{x_i\}$ are independent Brownian motions in the new ``time-like" variable $V$ with a diffusion constant  $D=1/2$. 
We compute the survival probability up to a fixed renormalized ``time" $V$ and then average over $V$ drawn from $h(V,t)$ in~\eref{eq:hV2}.

\subsection{Survival probability and first-passage time for a single particle}
\label{s:FPP}

It is useful to start with how the first-passage properties can be calculated for a single Brownian particle with a stochastic diffusion coefficient. The goal is to compute the first-passage probability $q_1(t;x)\, dt$ that a particle, starting from the origin, hits a target located at $x$, for the first time (hence the name first-passage time) between time $t$ and $t+dt$. The first-passage time PDF $q_1(t;x)$ is closely related to another standard quantity, the survival probability $W_1(t;x)$. The survival probability is defined as the probability that a particle, starting from the origin, has not reached (or crossed) the target located at $x$ up to time $t$. These two quantities are connected through  
\begin{equation}
W_1(t;x) = \int_t^\infty q_1(t';x)\,dt' ,
\end{equation}
which simply states that the probability of surviving until time $t$ equals the probability of first hitting the target at some later time $t'>t$. Equivalently, the first-passage time density can be obtained as  
\begin{equation}
q_1(t;x) = -\frac{\partial}{\partial t} W_1(t;x).
\label{eq:FPT-PDF}
\end{equation}
The survival probability satisfies the boundary conditions
\begin{equation}
W_1(0;x) = 1, \qquad W_1(\infty;x) = 0.
\end{equation}

Since, for a given realization of $\{D(\tau); 0 \leq \tau \leq t\}$, there is a one-to-one correspondence between $t$ and $V$ through~\eref{eq:V1}, it is convenient to first compute the survival probability as a function of $V$. For a fixed $V$, the particle positions evolve as a standard Brownian motion in the variable $V$ [see \eref{eq:LE_transformed}] with diffusion constant $D=1/2$. In this case, the survival probability up to the ``time'' $V$ is well known and is given by~\cite{Redner_2001}
\begin{equation}
    S(V;x) = \mathrm{erf}\!\left( \frac{|x|}{\sqrt{2 V}} \right) \, .
    \label{eq:conditional_survival}
\end{equation}
Averaging over the random variable $V$ with respect to its PDF $h(V,t)$, then yields the survival probability $W_1(t;x)$ for a single particle, 
\begin{equation}
    W_1(t;x) = \int_0^\infty dV \, h(V,t) \, S(V;x) \, .
    \label{eq:survival_single_pt_1}
\end{equation}
For the case, $D(t)=B^2(t)$, using the the explicit forms of $h(V,t)$ from \eref{eq:hV2} and \eref{Q(z)}, 
the integral in~\eref{eq:survival_single_pt_1} can be performed exactly, yielding
\begin{equation}
    W_1(t;x) = \frac{2 \sqrt{2}}{\pi} \sum_n \binom{-1/2}{n} \, \tan^{-1}{\left(\frac{|x|}{(4n+1) \Lambda t} \right)} \, .
    \label{eq:survival_single_pt2}
\end{equation}
Accordingly, the PDF of the first-passage time, using \eref{eq:FPT-PDF}, is given by 
\begin{equation}
    q_1(t;x) = \frac{\Lambda}{|x|} \, u\left( \frac{\Lambda t}{|x|} \right) \, ,
    \label{eq:FPT_scaling_form}
\end{equation}
where the scaling function
\begin{equation}
    u(z) = \frac{2 \sqrt{2}}{\pi} \sum_{n=0}^\infty  \, \binom{-1/2}{n} \frac{(4n+1)}{1 + (4n+1)^2 z^2}   \, .
    \label{eq:u(z)}
\end{equation}
Comparing \eref{eq:u(z)} with \eref{eq:fy2} and then using \eref{single particle distribution}, it is easy to see that
\begin{equation}
\label{eq:uz}
    u(z) = \frac{1}{2 z^2} f\left(\frac{1}{4z}\right),
\end{equation}
where the function $f(y)$ is given by~\eref{single particle distribution}. 
This result was obtained earlier in different ways~\cite{target_search, SBS2021, Grebenkov_2021}. However, here, our motivation was to introduce a method that can be easily generalized to many-particle systems. 

\begin{figure}[t]
    \centering    \includegraphics[width=0.55\textwidth]{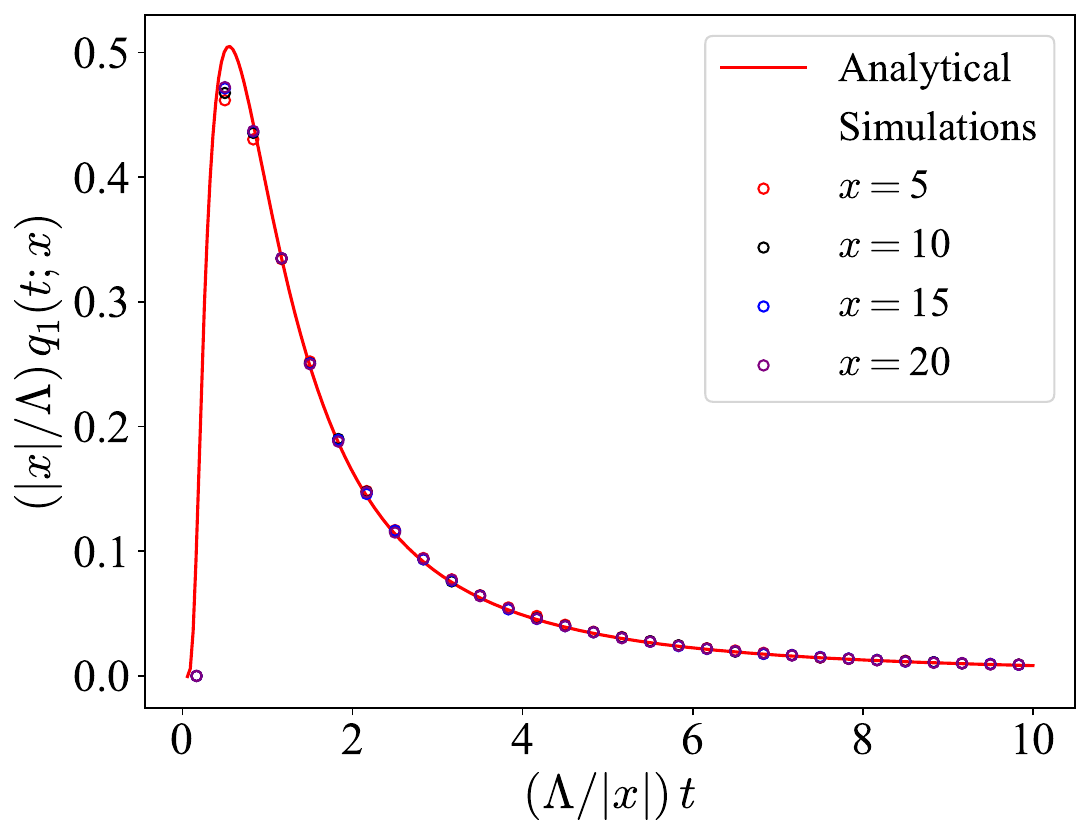}
    \caption{The PDF of the scaled first-passage time $(\Lambda/|x| ) \, t$, to a target at $x$, by a single particle starting from the origin. The red solid line plots the scaling function $u(z)$ in~\eref{eq:u(z)}. The points are obtained from simulation, with  $\Lambda=1$, $dt =0.01$, averaged over $10^6$ realizations.} 

    \label{fig:FPT}
\end{figure}

In the limit of large $z$, the dominant contribution comes from the $n=0$ term of the sum in~\eqref{eq:u(z)}. Consequently, the tail decays as $q_1(t;x)\sim t^{-2}$ for $t \gg x/\Lambda$, in contrast to the $t^{-3/2}$ tail of the standard Brownian motion with a constant diffusivity. Nevertheless, the mean first-passage time remains divergent, as in the case of the standard Brownian motion. 
In~\fref{fig:FPT}, we compare the scaling function in~\eqref{eq:u(z)} with the scaled collapsed simulation data for various $x$ and find excellent agreement.

Note that the leading power-law tail of $q_1(t;x)$ for large $t$ can be easily extracted. 
From~\eref{eq:conditional_survival}, we have $S(V;x)\sim V^{-1/2}$  for large $V$.
For $D(t)=B^2(t)$, since $B(t)\sim \sqrt{t}$, from~\eref{eq:V1} $V\sim t^2$, and consequently, from \eref{eq:conditional_survival} and \eref{eq:FPT-PDF}, we get $W_1(t;x)\sim t^{-1}$ and $q_1(t;x)\sim t^{-2}$, respectively.   

\begin{figure}
    \centering    
    \includegraphics[width=1\textwidth]{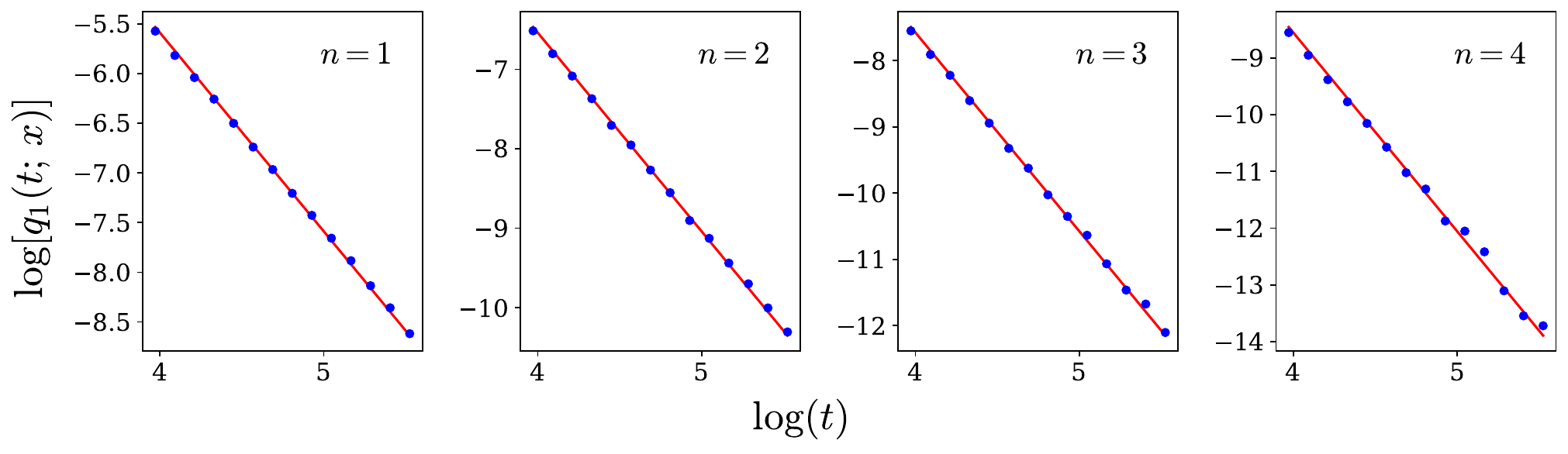}
    \caption{The tail of the PDF of the first-passage time in log-scale,  $\log[q_1(t;x)]$ vs $\log(t)$, for large $t$, for 
     the stochastic diffusivity $D(t) = B^{2n}(t)$ for various $n$. The solid line plots the function $y = (n+3) x/2 + c$, where the intercept $c$ is evaluated using the least squares method. The blue markers are obtained by direct numerical simulations described in \sref{sec:simulation}, with parameters $dt=0.01$, $x=10$, averaged over $10^6$ realizations. 
    } 

    \label{fig:FPT_DD}
\end{figure}

In general, for the class of diffusing diffusivity models defined by $D_n(t) = B^{2n}(t)$ with $n$ being a positive integer, we have $V\sim t^{n+1}$. Consequently, $W_1(t;x) \sim [V(t)]^{-1/2}\sim t^{-(n+1)/2}$ and $q_1(t;x) \sim t^{-(n+3)/2}$. We verify this power-law tail of $q_1(t;x)$ in  \fref{fig:FPT_DD} by comparing with numerical simulations.
Interestingly, the mean first-passage time becomes finite for $n>1$. Another observation is that a similar power law tail would exist even for a deterministic diffusion coefficient $D(t) \sim t^n$.

\subsection{Survival probability and first-passage time for a many-particle system}
\label{sec:EFPT}

The survival probability of a single particle discussed in \sref{s:FPP} can be easily generalized to many particles. In this case, the survival probability of $N$ Brownian motions in $V$, all starting at the origin, is evidently given by $[S(V;x)]^N$. Therefore, averaging over $V$ with respect to its PDF $h(V,t)$, one finds that the survival probability in $t$, as
\begin{equation}
    W_N(t;x)= \int_0^\infty dV \, h(V,t) \, [S(V;x)]^N \, .
    \label{eq:survival_single_pt_N}
\end{equation}
Using $h(V,t)$ and $S(V;x)$ from \eref{eq:hV2} and \eref{eq:conditional_survival}, respectively, we find that $W_N(t;x)$ has a scaling form
\begin{equation}
    W_N(t;x)= \hat{W}_N\left(\frac{\Lambda t}{|x|}\right), \quad \text{where}\quad \hat{W}_N (z)=\frac{1}{4z^2} \int_0^\infty dy\, Q\left(\frac{y}{4z^2}\right)\, \left[\mathrm{erf}\left(\frac{1}{\sqrt{y}}\right)\right]^N.
    \label{eq:Wn-scal}
\end{equation}

For large $N$, using the asymptotic properties of the error function, it is easy to see that the function $\bigl[\mathrm{erf}(1/\sqrt{y})\bigr]^N$ decays from unity to zero very fast, around $y\sim 1/\ln N$, as $y$ increases from zero.
More precisely, for large $N$, the convergence
\begin{equation}
  \left[\mathrm{erf}\left(\frac{1}{\sqrt{y}}\right)\right]^N = \left[1-\mathrm{erfc}\left(\frac{1}{\sqrt{y}}\right)\right]^N  \to \exp\left[-N \,\mathrm{erfc}\left(\frac{1}{\sqrt{y}}\right)\right]
  \label{eq:er1}
\end{equation}
reminds us of the extreme value Gumbel distribution, discussed in~\sref{Section:EVS}. Therefore, setting $y^{-1}=a_N+b_N w$ in the last expression in~\eref{eq:er1}, with appropriate $N$-dependent coefficients $a_N$ and $b_N$, we expect 
\begin{equation}
 \lim_{N\to\infty}\, N \,\mathrm{erfc}\left(\sqrt{a_N+b_N w}\right)   = e^{-w}.
 \label{eq:erfcN}
\end{equation}
The coefficients $a_N$ and $b_N$ can be determined from the above equation as follows. Setting $w=0$ in \eref{eq:erfcN} and inverting we get
\begin{equation}
    a_N = \left[\mathrm{erfc}^{-1}(1/N)\right]^2 = \ln N -\frac{1}{2}\left(\ln \ln N + \ln \pi \right) + O(1/\ln N). 
    \label{eq:aN}
\end{equation}
Similarly, differentiating \eref{eq:erfcN} with respect to $w$ and then setting $w=0$, we get
\begin{equation}
    b_N = \frac{\sqrt{\pi \,a_N}\,  e^{a_N}}{N} = 1 + O(\ln \ln N/\ln N). 
    \label{eq:bN}
\end{equation}
The variable $y$ can be written in terms of $w$ as
\begin{equation}
    y= \frac{1}{a_N + b_N w} = \frac{1}{a_N} - \frac{b_N}{a_N^2} w + O(b_N/a_N^3).
\end{equation}
Therefore, $\bigl[\mathrm{erf}(1/\sqrt{y})\bigr]^N$ in~\eref{eq:Wn-scal}, as a function of increasing $y$, drops sharply from $1$ to $0$, around $y=a_N^{-1}\sim 1/\ln N$, within a scale of $\Delta y\sim b_N/a_N^2\sim 1/(\ln N)^2$. Consequently, we can replace $\bigl[\mathrm{erf}(1/\sqrt{y})\bigr]^N$ in~\eref{eq:Wn-scal} with a step function $\theta(a_N^{-1}-y)$, which yields,
\begin{equation}
     \hat{W}_N (z)=\frac{1}{4z^2} \int_0^{a_N^{-1}}
     dy\, Q\left(\frac{y}{4z^2}\right) = \int_0^{(4 a_N z^2)^{-1}} Q(y)\, dy\, .
     \label{eq:Wnsc}
\end{equation}
Using the expression of $Q(y)$ from \eref{Q(z)}, the integral in~\eref{eq:Wnsc} can be evaluated and $\hat{W}_N(z)$ can be written as the sum of incomplete gamma functions, which we do not write explicitly here. 

\begin{figure}
    \centering    
    \includegraphics[width=0.55\textwidth]{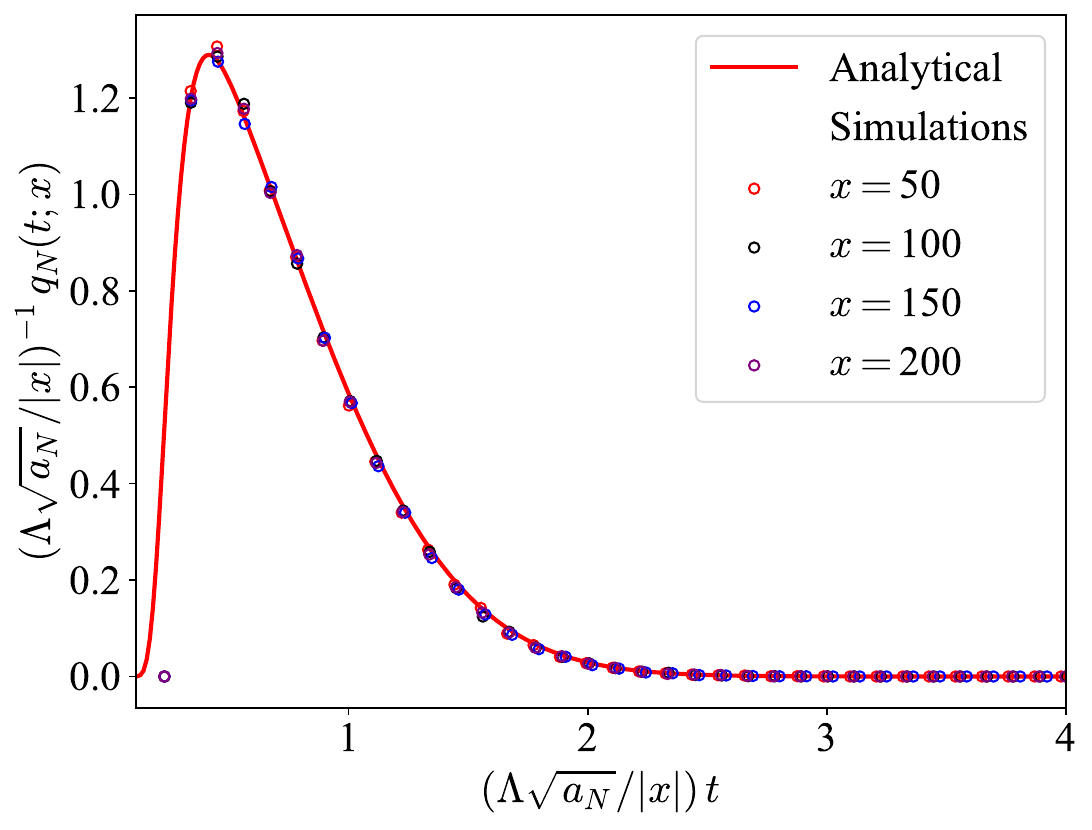}
    \caption{The PDF the scaled first-passage time $(\Lambda \sqrt{a_N}/|x|)\, t$, to a target $x$ by any one of the $N$ particles, all starting at the origin, where $a_N$ defined in Eq.~\eqref{eq:aN}. The red solid line plots the function of $[\Lambda \sqrt{a_N}]^{-1} q_N(t; x) \equiv U(z)$, with the scaling function $U(z)$ given in Eq.~\eqref{eq:U(z)}. The points are obtained from simulation with $\Lambda=1$, $N = 10^5$, $dt =0.1$, averaged over $10^5$ realizations.} 
    \label{fig:EFPT}
\end{figure}

The PDF $q_N(t;x)= -\partial_t W_N(t; x)$, of the first-passage time to the target at $x$, by any of the particles, can now be obtained by differentiating~\eref{eq:Wnsc} with respect to $z$. This gives the scaling form, 
\begin{equation}
    q_N(t;x)= \frac{\sqrt{a_N}\, \Lambda}{|x|}\, U\left(\frac{\sqrt{a_N}\, \Lambda}{|x|}\, t\right)\, ,
\end{equation}
where the scaling function $U(z)$ is given by
\begin{equation}
    U(z)=\frac{1}{2z^3}\, Q\left(\frac{1}{4z^2}\right)\, .
    \label{eq:FEsc}
\end{equation}
Using the explicit form of $Q(z)$ from~\eref{Q(z)}, we can write down the scaling function as
\begin{equation}
    U(z)=  \frac{2\sqrt{2}}{\sqrt{ \pi}} \, \sum_{n=0}^\infty \binom{-\frac{1}{2}}{n} \, (4n+1) \, \exp{\left[-(4n+1)^2\, z^2\right]} \, .
    \label{eq:U(z)}
\end{equation}
We compare this with numerical simulation in~\fref{fig:EFPT} and find excellent agreement. The leading tail behavior is given by the $n=0$ term of the summation in~\eref{eq:U(z)} as
\begin{equation}
    U(z) \sim \frac{2\sqrt{2}}{\sqrt{\pi}}\, e^{-z^2}, \quad\text{for large}~z.
\end{equation}
On the other hand, in the limit $z\to 0$, the summation in~\eref{eq:U(z)} tends to zero. More precisely, the limiting behavior of $Q(z)$ in~\eref{eq:Q-limiting} yields,
\begin{equation}
    U(z) \sim \frac{1}{2 z^3}\, e^{-\pi^2/(64 z^2)}\, , \quad\text{as }~ z\to 0.
\end{equation}
Therefore, the function $U(z)$ tends to zero as both ends $z\to 0$ and $z\to\infty$, and has a maximum in-between, as shown in~\fref{fig:EFPT}.

\subsection{First-exist time from an interval $[-L,L]$}

Another important temporal observable is the first-exit time from the region $[-L,L]$.
Like the first-passage time, the first-exit time is a random variable, and in this section, we want to calculate its PDF for the stochastic diffusivity $D(t) = B^2(t)$. To do this, we follow the same trick that was used in the previous sub-sections, i.e., first compute for a given $V$, and then average over $V$ in the second step.

In the first step, for a single Brownian motion in $V$ with a constant diffusion coefficient $D=1/2$, the probability $\Psi(V;  L)$ that the particle, starting from the origin,  has not escaped the region $[-L,L]$ up to ``time" $V$, is given by~\cite{MSS16}
\begin{equation}
\label{eq:sur_FET}
    \Psi(V; L) = 
    \frac{4}{\pi} \sum_{n=0}^\infty \frac{(-1)^n}{2n+1} \, \exp{\left[ - \frac{(2n+1)^2 \pi^2 V}{8 L^2} \right]} \, .
\end{equation}
In the second step,  averaging over $V$ with respect to its PDF $h(V,t)$
yields the survival probability in the interval $[-L,L]$ as
\begin{equation}
    \Phi(t; L) = \int_0^\infty dV \, h(V,t) \,\Psi(V;L)\, .
\end{equation}
Using the the explicit forms of $h(V,t)$ from \eref{eq:hV2} and \eref{Q(z)}, and $\Psi(V;x)$ from \eref{eq:sur_FET}, the above integral can be performed explicitly, which yields
\begin{equation}
    \Phi(t;  L) = \frac{4}{\pi} \sum_{n=0}^\infty \, \frac{(-1)^n}{2n+1} \, \sqrt{\mathrm{sech}\left[\frac{(2n +1) \pi\Lambda  \, t}{2L} \right]} \, .
\end{equation}
We have used \eqref{eq:LThV} to evaluate the averaging of Eq.~\eqref{eq:sur_FET} with $h(V,t)$. 

\begin{figure}
    \centering    
    \includegraphics[width=0.5\textwidth]{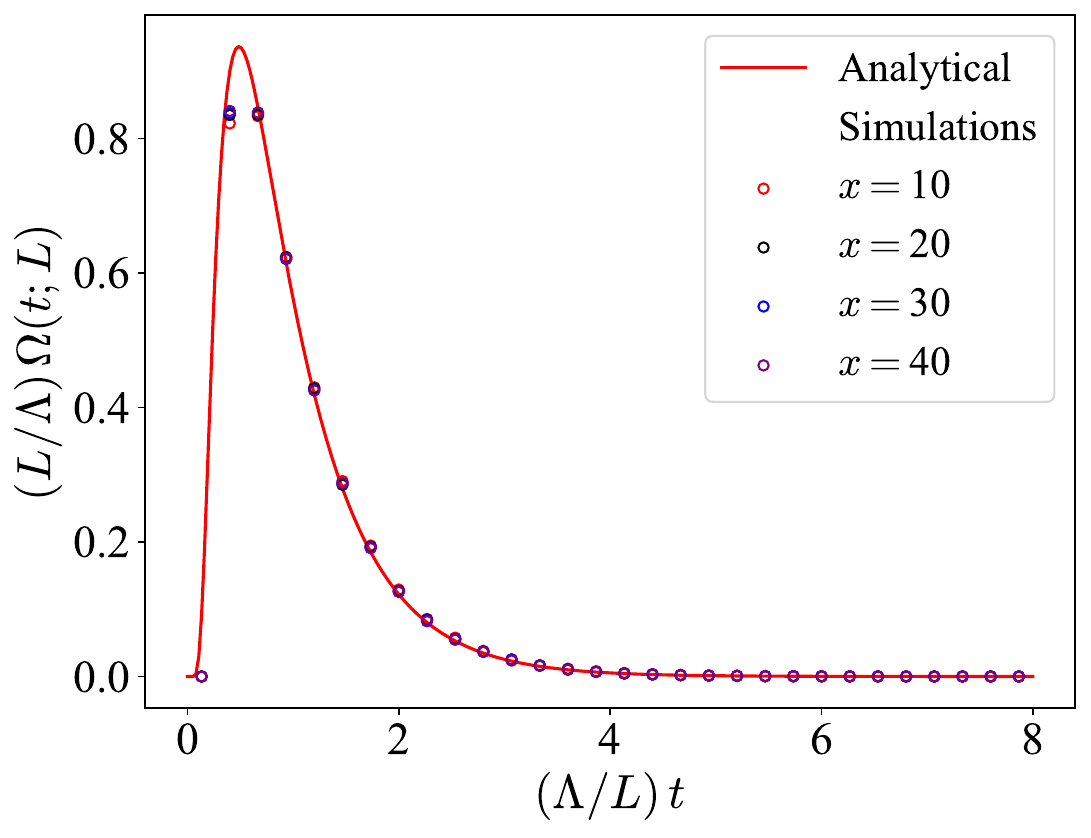}~\includegraphics[width=0.5\textwidth]{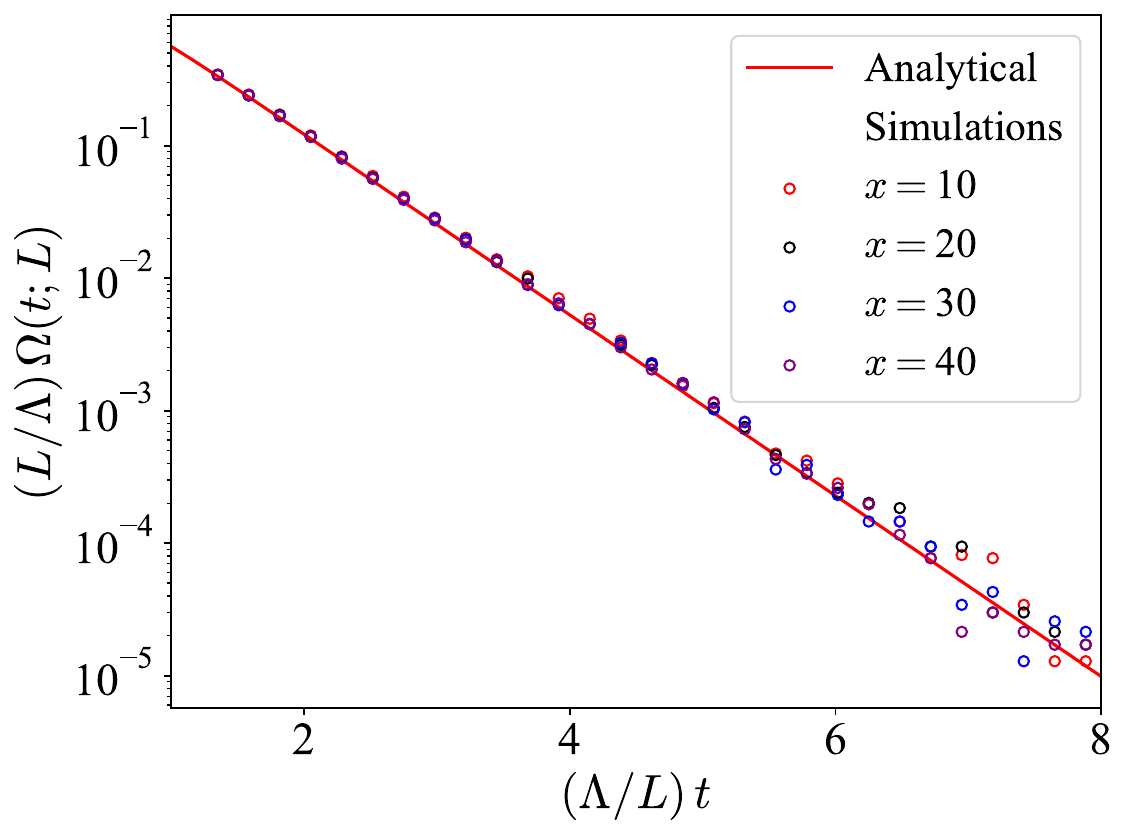}
\caption{The PDF of the scaled first-exit time $(\Lambda/L) \, t$
    from a region $[-L,L]$, by a single particle starting at the origin: on a linear-linear scale (left) and a log-linear scale (right).
    The red solid line is the plot of $(L/\Lambda) \Omega(t; \, L) \equiv \chi(z)$, with the scaling function $\chi(z)$ given in~\eref{eq:FET_scale_fn}. The markers indicate simulation performed with parameters $\Lambda=1$, $dt =0.01$, averaged over $10^6$ realizations.} 

    \label{fig:FET}
\end{figure}

The PDF of the first exit time, given by $\Omega (t;  L) = - \partial_t \Phi(t  ;L)$, can be evaluated as
\begin{equation}
    \Omega(t ;  L) = \frac{\Lambda}{L} \, \chi\left(\frac{\Lambda \, t}{L}\right) \, ,
\end{equation}
where the scaling function is given by 
\begin{equation}
\label{eq:FET_scale_fn}
    \chi(z) = 2  \sum_{n=0}^{\infty} \, (-1)^n \, \sqrt{\mathrm{sech}{\bigl[(2n +1) \pi z\bigr]}}\, \,  \tanh\bigl[(2n +1) \pi z\bigr]\, .
\end{equation}
At long times, the PDF $\Omega(t;  L)$ decays exponentially $\Omega(t; L) \sim (\Lambda t/L) e^{- \pi \Lambda t/(4L)}$, just like the case of conventional Brownian motion. However, unlike the case of conventional Brownian motion, here the PDF of the first exit time scales with $L$ rather than $L^2$. We compare the scaling function in~\eref{eq:FET_scale_fn} in~\fref{fig:FET} with numerical simulation and find excellent agreement.  

We can now generalize easily to $N$ particles, all starting with the origin. In this case, as before, the survival probability in the interval $[-L,L]$ is given by
\begin{equation}
\label{eq:int_Phi_N}
    \Phi_N(t; L) = \int_0^\infty dV \, h(V,t) \,[\Psi(V;L)]^N\,  \,  .
\end{equation}
To determine the large $N$ behavior of $\left[ \Psi(V \, ;L)\right]^{N}$, we need to analyze the small $V/L^2$ behavior of the function $\left[ \Psi(V \, ;L)\right]$, and  \eref{eq:sur_FET} is not suitable for this.   Instead, we need to use the Poisson summation formula, which, to the leading order, gives~\cite{MSS16}
\begin{equation}
    \Psi(V;L) \simeq 1-2\,\mathrm{erfc}\left(\frac{L}{\sqrt{2V}}\right).
\end{equation}
Therefore, 
\begin{equation}
 [\Psi(V;L)]^N \to  \exp\left[-2N \,\mathrm{erfc}\left(\frac{L}{\sqrt{2V}}\right)\right] \quad\text{for large } N\, .
 \label{eq:psiVN}
\end{equation}
This expression is indeed similar to that in~\eref{eq:er1}. Therefore, 
\begin{equation}
    \left[\Psi\left(V=\frac{1}{\tilde{a}_{N,L}} - \frac{\tilde{b}_{N,L}}{\tilde{a}_{N,L}^2} w;L\right)\right]^N \to \exp\left(-e^{-w}\right)\quad\text{as }~~N\to\infty,
\end{equation}
where 
\begin{equation}
    \tilde{a}_{N,L} = \frac{2\,a_{2N}}{L^2}\quad\text{and}\quad \tilde{b}_{N,L} = \frac{2\, b_{2N}}{L^2}\, ,
\end{equation}
with $a_N$ and $b_N$ given in \eref{eq:aN} and \eref{eq:bN} respectively. This implies that $[\Psi(V;L)]^N$, as a function of increasing $V$, drops sharply from $1$ to $0$ around $V=a_{N,L}^{-1}\sim L^2/[2\ln N]$ within a scale of $\Delta V \sim L^2/[2 (\ln N)^2]$. Consequently, \eref{eq:int_Phi_N} can be well-approximated by
\begin{equation}
    \Phi_N(t;L)= \int_0^{a_{N,L}^{-1}} h(V,t)\, dV\, .
\end{equation}
Finally, using the scaling form \eref{eq:hV2} for $h(V,t)$, we get 
\begin{equation}
    \Phi_N(t;L)=\hat{\Phi} \left(\frac{\sqrt{a_{2N}}\, \Lambda\, t}{L}\right), \quad\text{where}\quad
\hat{\Phi}(z) = \int_0^{1/(4z^2)}\, Q(y)\, dy\, .
\end{equation}

\begin{figure}
    \centering    
    \includegraphics[width=0.55\textwidth]{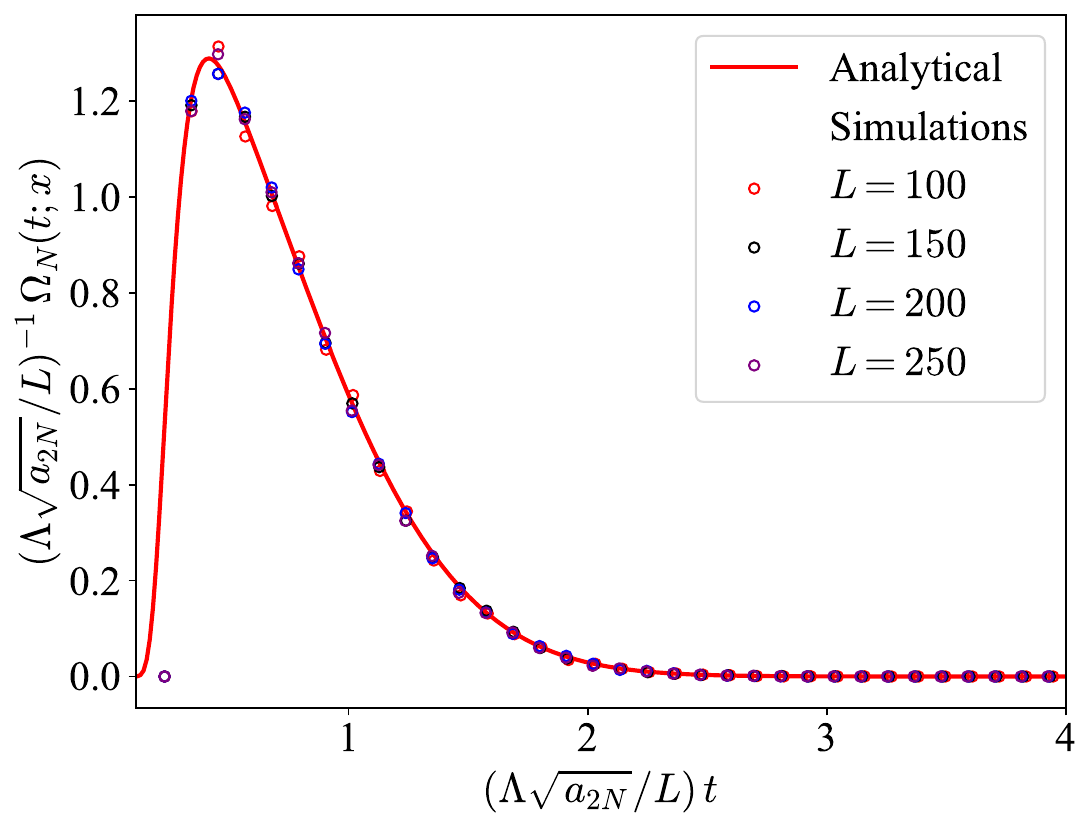}
    \caption{The PDF of the scaled first-exit time $(\Lambda \sqrt{a_{2N}}/L)\, t$,
    from the region $[-L,L]$, by any one of the $N$ particles, all starting at the origin, where $a_N$  is defined in~\eref{eq:aN}. The red solid line is the plot of $[\Lambda \sqrt{a_{2N}}/L]^{-1} \Omega_N(t; L) \equiv U(z)$, with the scaling function $U(z)$ given in~\eref{eq:PDFNFE}, with its explicit form in~\eref{eq:U(z)}. The markers indicate simulation performed with parameters $\Lambda=2$, $N = 10^5$, $dt =0.1$, averaged over $10^5$ realizations.} 
    \label{fig:EFET}
\end{figure}

The PDF $\Omega_N(t;L)=-\partial_t \, \Phi_N(t;L)$ of the first-exit time from the box $[-L,L]$ by any one of the $N$ particles, admits the scaling form 
\begin{equation}
    \Omega_N(t;L) = \frac{\sqrt{a_{2N}}\, \Lambda}{L} \, U\left(\frac{\sqrt{a_{2N}}\, \Lambda\, t}{L}\right)\, ,
\end{equation}
where the scaling function  
\begin{equation}
    U(z)= - \hat{\Phi}'(z) = \frac{1}{2 z^3}\, Q\left(\frac{1}{4z^2}\right)\,,
    \label{eq:PDFNFE}
\end{equation}
is the same
as that of the first-passage time
in~\eref{eq:FEsc}. 
Interestingly, the PDF of the first-exit time from a box $[-L,L]$ by $N$ particles is identical to that of the first-passage time to a target at $L$ by $2N$ particles. In~\fref{fig:EFET} we compare~\eref{eq:PDFNFE}, using its explicit form in~\eref{eq:U(z)}, with numerical simulation and find excellent agreement.

\section{Details of numerical simulation}
\label{sec:simulation}

In this section, we provide some details of the numerical simulation used in this paper.

\subsection{Simulations for $h(V,t)$}
\label{sec:sim_h_V}

To compute the PDF $h(V, t)$ of the random variable $V$ in~\eref{eq:defV}, we discretize the evolution equation~\eref{eq:bm} for  
Brownian motion $B(t)$,  in time steps of $dt$,  
\begin{equation}
    B(t_{n+1}) = B(t_n) + \sqrt{2 \Lambda^2 dt} \,\, r_n \, , \ \text{with }B(0)=0,  
    \label{eq:sim_B_evolv}
\end{equation}
where $t_n = n \, dt$. Here, $\{r_n\}$ are independent Gaussian random numbers with zero mean and unit variance.
The integral in Eq.~\eqref{eq:defV} is then performed numerically to evaluate $V$ for a given realization,
\begin{equation}
    V(t) = 2 \, dt \, \sum_{n=0}^{N_t} B^2(t_n) \, , 
    \label{eq:V-sim}
\end{equation}
with  $t = N_t dt$. The distribution of $V$ is estimated by repeating its evaluation using~\eref{eq:V-sim} over $10^8$ independent realizations and constructing a histogram with bin width $dV$. Dividing the bin counts by $dV$ then yields the PDF $h(V,t)$, which is plotted in figure~\ref{fig:h(V)}.

\subsection{Simulations for various observables $\mathcal{O}(x_1, x_2, \dotsc, x_N)$}
All our observables discussed in this article are of the form $\mathcal{O}(x_1, x_2, \dotsc, x_N)$, which depend on the position coordinates $\{x_1(t), x_2 (t), \dotsc, x_N(t)\}$. These positions evolve according to the Langevin equation~\eref{eq:LE}, where the diffusion coefficient $D(t)=B^2(t)$. Therefore, we discretize~\eref{eq:LE} in time steps of $dt$ as \begin{equation}
    x_i(t_{n+1}) = x_i(t_n) + \sqrt{2 B^2(t_n) \, dt} \,\, p_{i,n} \, ,
    \label{eq:LE-disc}
\end{equation}
where $p_{i,n}$ are independent Gaussian random numbers (for both indices $i,n$) with mean zero and unit variance. The Brownian motion $B(t_n)$ in~\eref{eq:LE-disc} is evolved at the same time by~\eref{eq:sim_B_evolv} in time steps of $dt$. Therefore, at any given time $t_n$, we can compute the observable $O(x_1(t_n),x_2(t_n), \dotsc, x_N(t_n))$, for any given realization of the set of independent random variables $\{r_n, p_{i,n}\}$. We can then obtain various moments of $\mathcal{O}$ or its distribution by averaging over a large number of realizations, typically $10^5-10^6$.

\section{Conclusion}
\label{s:conclusion}
 
 In this paper, we have studied a gas of $N$ Brownian particles subjected to a common stochastic diffusivity $D(t)=B^2(t)$, where $B(t)$ is another independent Brownian motion with a constant diffusion coefficient. We start with the initial condition where the gas is localized at the origin. The gas expands ballistically over time with a length scale $\xi(t)\propto t$, and we obtained the exact joint probability density function (JPDF) of the position of all the particles at all times $t$. We have shown that there is a dynamically generated all-to-all attraction between particles at any finite $t>0$, even though there is no direct interaction between the particles. Moreover, we have shown that the JPDF at any finite time $t$, though not factorizable into individual marginal distributions, indicating strong non-zero correlations, has a 
 special conditionally independent and identically distributed (CIID) structure. This CIID structure allows us to compute several macroscopic and microscopic observables exactly at all times $t$, despite the presence of strong correlations at finite time $t>0$. These observables include the average density profile of the gas, the extreme and the order statistics of the positions of the particles, the distribution of the spacing between two consecutive particles, and also the full counting statistics, i.e., the distribution of the number of particles in an interval $[-L, L]$. They exhibit a rich variety of interesting behaviors.  We also obtain the PDFs of the first-passage time to a target at $x$ and the first-exit time from a box $[-L,L]$ by any one of the particles, all starting at the origin. Interestingly, the PDF of the first-exit time
 from a box $[-L,L]$ for $N$ particles is identical to that of the first-passage time
 to a target at $x=L$ by $2N$ particles, for large $N$.
 
 In this paper, we have used the choice of the fluctuating diffusivity $D(t)=B^2(t)$. It would be interesting to study other choices of stochastic diffusivity, such as the square of an Ornstein-Uhlenbeck process. It would also be interesting to study such dynamically correlated gas in a confining potential. Finally, here we have considered a gas where there are no direct interactions between the particles. It would be interesting to extend this study, where there is already a direct interaction between the particles, and then subject the system to a common fluctuating environment, as in some of the recent studies~\cite{DMS2025, biroli2025resetting}.

\section{Acknowledgements}
 SNM and SS acknowledge the support from the Science and Engineering Research Board (SERB, Government of India), under the VAJRA faculty scheme No. VJR/2017/000110. SNM acknowledges the ANR Grant No. ANR- 23- CE30-0020-01 EDIPS. 
  SNM and SS thank the support from the International Research Project (IRP) titled `Classical and quantum dynamics in out of equilibrium systems' by CNRS, France.
 SS thanks the hospitality of Laboratoire de Physique Th\'eorique et Mod\`eles Statistiques (LPTMS)
and Laboratoire de Physique Th\'eorique et Hautes Energies (LPTHE), Sorbonne Universit\'e, Paris, France. 

\vskip0.5cm

\section*{References}


\providecommand{\newblock}{}

\end{document}